\documentclass[final,1p,times,authoryear]{elsarticle}
\usepackage{tikz}
\usetikzlibrary{patterns}
\usepackage{booktabs} % For better table formatting

\newcommand{\blackline}{\raisebox{2pt}{\tikz{\draw[-,black!40!black,solid,line width = 0.9pt](0,0) -- (5mm,0);}}}

\newcommand{\blueline}{\raisebox{2pt}{\tikz{\draw[-,blue!40!blue,solid,line width = 0.9pt](0,0) -- (5mm,0);}}}

\newcommand{\redline}{\raisebox{2pt}{\tikz{\draw[-,red!40!red,solid,line width = 0.9pt](0,0) -- (5mm,0);}}}

\usepackage{tikz}
\usepackage{xcolor}
\emergencystretch=\maxdimen
\definecolor{matplotbrown}{RGB}{139, 69, 19}
\definecolor{matplotpink}{RGB}{255, 20, 147}

\definecolor{matplotcyan}{RGB}{0, 191, 255}

\newcommand{\greenline}{\raisebox{2pt}{\tikz{\draw[-,green!40!green,solid,line width = 0.9pt](0,0) -- (5mm,0);}}}

\usepackage{graphicx}%
\usepackage{multirow}%
\usepackage{multicol}
\usepackage{amsmath,amssymb,amsfonts}%
\usepackage{amsthm}%
\usepackage{mathrsfs}%
\usepackage[title]{appendix}%
\usepackage{xcolor}%
\usepackage{textcomp}%
\usepackage{manyfoot}%
\usepackage{algorithm}%
\usepackage{algorithmicx}%
\usepackage{algpseudocode}%
\usepackage{listings}%
\usepackage[labelfont=it]{subcaption}
\usepackage{hyperref}
\usepackage[process=all]{pstool}
\theoremstyle{thmstyleone}%
\theoremstyle{thmstyletwo}%

\theoremstyle{thmstylethree}%

\makeatletter
\providecommand\HyPL@Entry[1]{}
\AddToHook{env/document/begin}{%
\@ifpackageloaded{preview}{
\ifPreview
 \let\Hy@FirstPageHook\relax
 \let\Hy@EveryPageAnchor\relax
\fi}{}}
\makeatother
\journal{International Journal of Heat and Fluid Flow}

\begin{document}

\begin{frontmatter}



\title{To stall-cell or not to stall-cell: Variational data assimilation of 3D mean flow past a stalled airfoil} %% Article title


\author[inst1]{Uttam Cadambi Padmanaban} %% Author name
\ead{ucp1n22@soton.ac.uk}

\affiliation[inst1]{organization={Aeronautics and Astronautics, University of Southampton},%Department and Organization
            addressline={University Road}, 
            city={Southampton},
            postcode={SO17 1BJ},
            country={United Kingdom}}

\author[inst2]{Craig Thompson} %% Author name
\ead{craig.thompson@imperial.ac.uk}
\affiliation[inst2]{organization={Aeronautics and Astronautics, Imperial College London},%Department and Organization
            addressline={Exhibition Rd, South Kensington}, 
            city={London},
            postcode={SW7 2AZ},
            country={United Kingdom}}

\author[inst1]{Bharathram Ganapathisubramani} %% Author name
\ead{g.bharath@soton.ac.uk}

\author[inst1]{Sean Symon} %% Author name
\ead{sean.symon@soton.ac.uk}

%% Abstract
\begin{abstract}
The full-field reconstruction of three-dimensional (3D) turbulent flows from sparse experimental measurements remains a significant challenge, particularly for flows exhibiting complex 3D flow separation. In this work, we address this challenge for the case of stall cells - spanwise coherent structures that form on the suction surface of wings at post-stall conditions. Planar particle image velocimetry (PIV) experiments are performed on a NACA 0012 wing at a chord-based Reynolds number of $Re_c \approx 450{,}000$ and angle of attack $\alpha = 14^\circ$, acquiring two-component mean velocity data on four spanwise planes. The experimental data show clear spanwise variation in the extent of the separation and flow dynamics, consistent with the presence of stall cells. Three-dimensional variational (3DVar) data assimilation (DA) within the field inversion framework is then employed to reconstruct the full 3D mean flow field by augmenting these sparse planar measurements with the Spalart--Allmaras (SA) Reynolds-averaged Navier--Stokes (RANS) turbulence model. The performance of the reconstruction is assessed on planes not used in the assimilation. It is shown that a single plane of sparse experimental data is sufficient to recover the essential features of a stall cell, including counter-rotating vortices around focal points on the suction surface. The lowest reconstruction error is obtained when two planes of data that are close together but exhibit markedly different separation extents are used, and the complementary roles of the reference data placement and the computational boundary conditions in shaping the reconstructed stall cell structure are explained. These results demonstrate the capability of 3DVar DA to reconstruct the full 3D physics of stall cells from two-component velocity data acquired on select spanwise planes.
\end{abstract}  

\begin{graphicalabstract}
\includegraphics[scale=0.3]{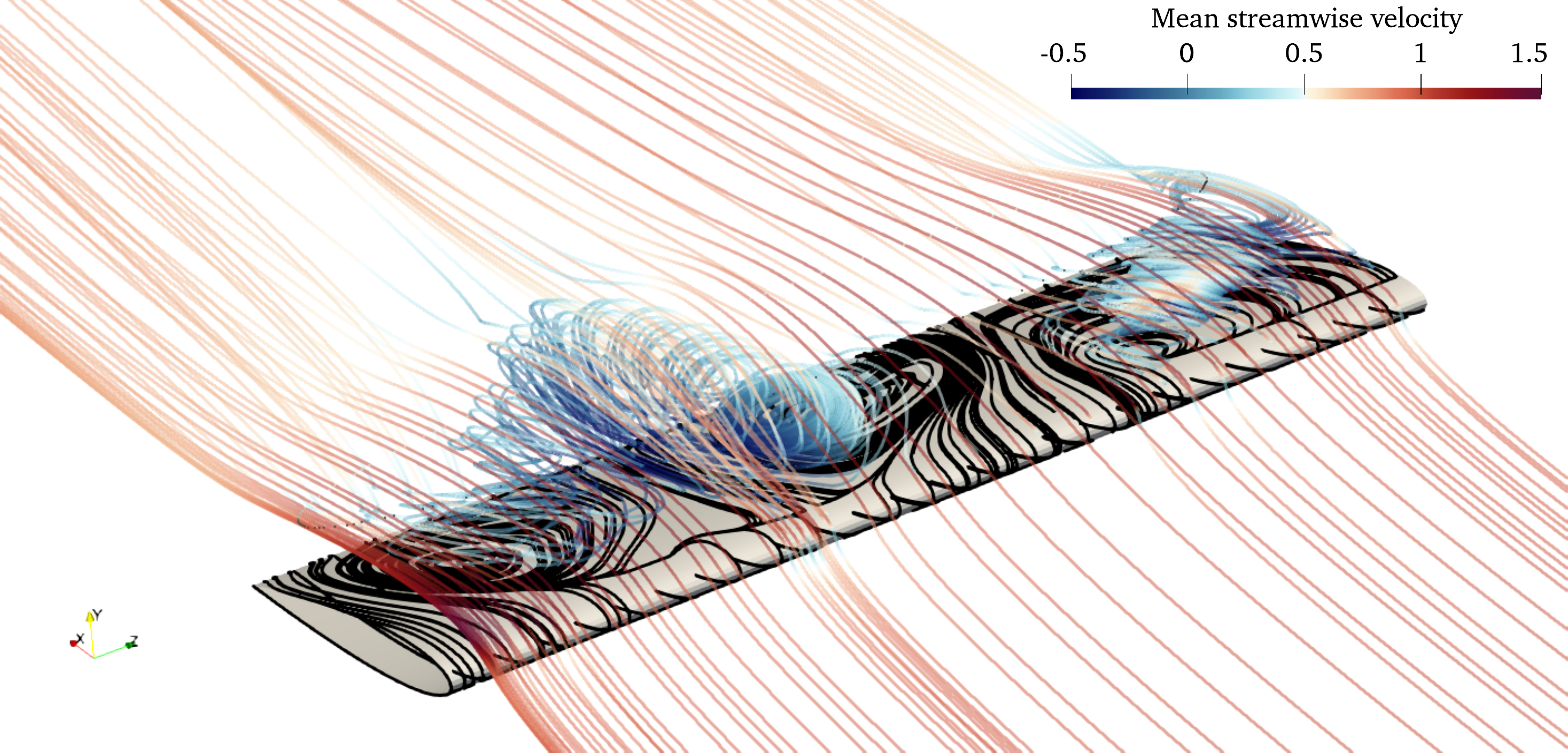}
\end{graphicalabstract}

\begin{highlights}
\item Three-dimensional stall cell structures on the wing surface are reconstructed from sparse two-component planar particle image velocimetry (PIV) data on a NACA 0012 wing at $Re_c \approx 450{,}000$ and $\alpha = 14^\circ$ using variational data assimilation within the field inversion framework.
\item A single plane of experimental data is sufficient to recover the essential features of a stall cell, including counter-rotating vortices around focal points on the suction surface of the wing.
\item Two planes of data that are close together but exhibit markedly different mean separation extents produce the most compact and well-defined stall cell reconstruction.
\item The reconstructed stall cell topology is robust to variations in the number and placement of reference data planes, with the position of the reconstructed focal point of the vortex that is closest to the computational domain boundary being consistent across all cases.
\end{highlights}

%% Keywords

\end{frontmatter}



\section{Introduction}
\label{section: Introduction}
The experimental study of turbulent flows relies on field imaging techniques such as particle image velocimetry \citep[PIV;][]{adrian1991particle} that provide access to two-component (planar) or three-component (stereo) mean velocity fields along a single plane. The imaging of three-dimensional (3D) flows requires such measurements to be carried out on multiple planes, immediately increasing the complexity of the optical setup, storage requirements, and the time required to perform the experiments and process the data. An alternative is to use tomographic PIV \citep{elsinga2006tomographic}, which provides volumetric velocity data, but it is considerably more complex and less widely available. As such, full-field reconstruction of turbulent flows from experimental data remains a significant challenge. 

There are other challenges associated with experimental measurements. The mean velocity data obtained using PIV are often sparse and have a limited field of view, capturing only a small portion of the flow volume. This is compounded by measurement uncertainties \citep{sciacchitano2019uncertainty} and the absence of a divergence-free velocity field when only two components are measured \citep{de2013minimization}. These challenges are particularly acute for 3D separated flows, where the spanwise organization of the flow cannot be captured from a single measurement plane. Scale-resolving computational fluid dynamics (CFD) approaches such as direct numerical simulation (DNS) and large eddy simulation (LES) can provide full 3D flow fields, but their cost scales unfavorably with the Reynolds number \citep{pope2001turbulent, choi2012grid}, making them prohibitive for flows of practical relevance. Reynolds--averaged Navier–Stokes (RANS) simulations offer a computationally tractable alternative by modeling turbulence through a closure relation, but they have limited accuracy for separated flows. Obtaining an accurate mean flow field is important not only for studying the flow physics but also as a prerequisite for stability and resolvent analysis \citep{symon2019tale, sarras2024linear,von2024role}. These considerations motivate the development of approaches that combine the computational efficiency of RANS with the accuracy of experimental measurements.

Among the 3D separated flows where full-field reconstruction is especially desirable are those exhibiting stall cells. These are spanwise coherent structures that form on the suction surface of wings at post-stall angles of attack. Stall cells were first documented as 3D separation patterns on rectangular wings by \cite{moss1971two} and \cite{gregory1971progress}, and further characterized as ``mushroom-shaped'' structures by \cite{winkleman1982}. Flow-field models were subsequently proposed to describe their vortex topology \citep{winklemann1980, weihs1983cellular}. Tuft and pressure measurements provided evidence for their dynamic nature, characterized by spanwise movement and jostling \citep{yon1998study}. \cite{manolesos2014experimental} provided a systematic characterization of stall cells on a rectangular wing, documenting the dependence of their geometry on Reynolds number, angle of attack, and aspect ratio (AR), while more recent 3D time-resolved measurements have characterized their physics in significantly greater detail \citep{manolesos2014study, dell2016measurement, neal2023three}. There has also been an effort to provide a more unified indicator for the emergence of stall cells using the lift polar \citep{gross2015criterion, de2022effects}. While experimental techniques have advanced significantly, the full 3D structure of stall cells remains difficult to characterize from planar experimental measurements alone, motivating the use of computational methods to complement the experimental observations.

In the specific context of stall cells, the use of DNS or LES is prohibitively expensive due to the need to simulate high aspect ratio wings with sufficient grid discretization along the span \citep{manni2016numerical, liu2018numerical}. RANS simulations offer a computationally tractable alternative but their performance for stall cell prediction is limited. Periodic boundary conditions require a sufficiently large AR domain to accommodate the spanwise wavelength \citep{manni2016numerical}. Symmetry boundary conditions have also been employed \citep{gross2010numerical}. However, quantitative agreement with experiments remains poor regardless of the boundary treatment. \cite{manolesos2014experimental} found a consistent delay of approximately $3^\circ$ in the predicted angle of attack for which stall cells appear using the Spalart--Allmaras (SA) model, and \cite{sarras2024linear} showed that the baseline SA model predicted the appearance of stall cells at $\alpha \approx 15.5^\circ$ compared to an experimental value of $\alpha \approx 12^\circ$ for a NACA 4412 at $Re_c = 350{,}000$. The inability of RANS to accurately predict the onset and structure of stall cells, combined with the expense of scale-resolving simulations, makes this an ideal candidate for data-driven correction of the turbulence closure using experimental observations.

Data assimilation (DA) provides a framework for combining model predictions with observational data to produce corrected flow fields that extend beyond the spatial extent of the original measurements. Originating in numerical weather prediction \citep{le1986variational}, variational DA has been widely adopted in the fluid mechanics community. In this formulation, the discrepancy between model output and observational data is minimized by tuning a control variable, and it has been applied extensively over the past decade to problems ranging from canonical flows \citep{foures2014data, franceschini2020mean, brenner2022efficient} to separated flows over airfoils \citep{symon2019tale, mons2024data, cadambi2026three}. The control variable can be placed at various locations in the governing equations, with corrections to the turbulence transport equations being particularly relevant for separated flows \citep{thompson2024effect, cadambi2024towards, mons2024data}. The field inversion framework introduced by \cite{singh2017augmentation} uses a scalar multiplier $\beta$ on the production term of the SA turbulence model, which is optimized using high-fidelity data to bring the RANS solution into agreement with the reference observations. This also constitutes the first step of the field inversion and machine learning (FIML) pipeline, which enables the generalization of optimized corrections to unseen flow configurations \citep{wu2025data}. Alternative data-driven approaches such as physics-informed neural networks (PINNs) have also been applied to mean flow reconstruction for canonical flows \citep{patel2024turbulence}, free-shear flows \citep{von2022mean}, and separated flows over airfoils \citep{christian2025review}. However, the majority of these studies have been limited to 2D configurations, often rely on high-fidelity reference data rather than experimental measurements, and have not been applied to the reconstruction of complex 3D flows using sparse experimental data.

Extending these approaches to 3D flows introduces additional challenges, and the existing body of work on 3D DA remains limited. Planar PIV measurements are inherently not divergence-free due to the absence of the out-of-plane velocity component \citep{cadambi2026three}, and the data requirements for 3D reconstruction can be considerable. \cite{steinfurth2024assimilating} required three-component mean velocity fields from stereo PIV in addition to surface pressure and skin friction measurements to reconstruct a turbulent separation bubble in a diffuser using PINNs. Within the variational framework, \cite{he2018data} employed the field inversion framework for several flows, including the 3D flow over a wall-mounted cube, using experimental velocity data sampled along discrete vertical profiles. \cite{cadambi2024towards} extended this to a single plane of three-component data for the same configuration using the FIML framework. \cite{cadambi2026three} addressed the divergence issue for a 3D flow past an airfoil by allowing the assimilated flow to develop the missing velocity component through the governing equations, though a full-field reconstruction was not the objective of that study. In the specific context of stall cells, \cite{sarras2024linear} performed DA using span-averaged DNS data at conditions matching an experiment where stall cells were observed, obtaining the corrected mean flow and eddy viscosity fields required for linear stability analysis. While this demonstrated the value of DA for stall cell prediction, the reliance on DNS as reference data limits the approach to configurations where such simulations are feasible.

The present study employs 3D variational (3DVar) DA within the field inversion framework \citep{singh2017augmentation} to reconstruct the full 3D flow field over a NACA 0012 wing at a chord-based Reynolds number of $Re_c \approx 450{,}000$ and an angle of attack of $\alpha = 14^\circ$. Planar PIV experiments are performed to image the flow along four spanwise planes. Two-component mean velocity data from these PIV measurements are used as input data to optimize a scalar multiplier of the production term in the turbulence transport equation, defined over the entire computational domain. The fully reconstructed flow field exhibits the features of stall cells, both in the bulk of the flow and on the surface, which are analyzed and compared to the existing body of experimental and computational evidence. In particular, we demonstrate the use of very sparse measurements (at most two planes of two-component data) and quantify the influence of the number and placement of PIV measurement planes on the reconstruction of the full flow field and the resulting surface flow topology. A full-field 3D reconstruction of stall cells from sparse planar experimental data, where the reconstruction is assessed on planes not used in the assimilation, has not yet been demonstrated. The performance of the reconstruction is assessed by comparing the assimilated flow to planes that are not used in the assimilation.

The remainder of this paper is organized as follows. Section \ref{section:mean_flow_experimental_data} describes the experimental setup and mean flow fields, including evidence for the presence of stall cells. Section \ref{section:turbulence_data_assimilation} outlines the turbulence modeling and DA methodology. Section \ref{section:case_setup_bsl_computation} presents the test case setup and results from the baseline SA model. Section \ref{section:stall_cell_reconstruction} discusses the results on assimilating single-plane and dual-plane mean velocity fields which include surface flow topology with concluding remarks in Section \ref{section:conclusion}.

\section{Experimental data of stall cells}
\label{section:mean_flow_experimental_data}
In this section, we provide details regarding the experimental setup and analyze the mean velocity fields. Section \ref{subsection:experimental_setup} presents the experimental setup for measuring aerodynamic forces and the mean velocity field using a load cell and PIV, respectively. Section \ref{subsection:experimental_mean_velocity_fields} presents the mean streamwise and wall-normal components of velocity along the wingspan at multiple spanwise locations. Section \ref{subsection:statistical_stationarity} provides evidence for the statistical stationarity of the mean velocity fields and the reproducibility of the stall cell position across experimental runs. Section \ref{subsection:POD} examines the spanwise coherence of the flow dynamics using proper orthogonal decomposition (POD). 

\subsection{Experimental setup}
\label{subsection:experimental_setup}

All experiments are carried out in the University of Southampton's “Gottingen” type closed-circuit wind tunnel. Each test section has an internal width of $1.2$~m, internal height of $1$~m and length of $2.4$~m. The facility includes a cooling system consisting of two heat exchangers and a temperature control unit, ensuring that the air temperature within the test section is maintained at a constant level. All experiments are run at a chord-based Reynolds number of $Re_c \approx 450,000$ at an angle of attack of $\alpha = 14^\circ$ with a free-stream turbulence intensity below $0.1$~\%.

\begin{figure*}[htbp]
\centering
\includegraphics[scale=0.25]{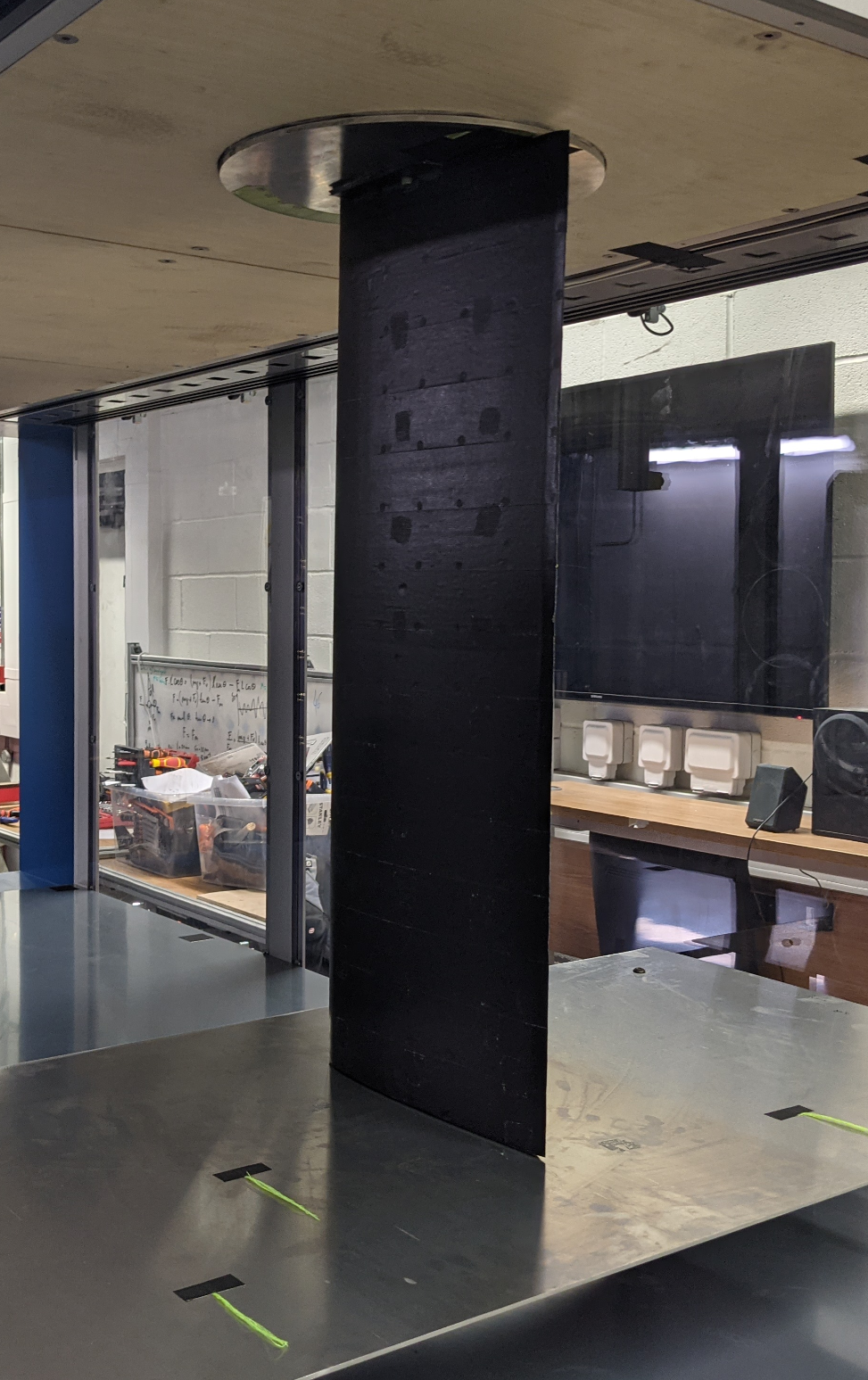}
\caption{Experimental wing setup.}
\label{fig:WingSetup}
\end{figure*}

The wing has a NACA 0012 profile with a $0.75$~m span and a $0.3$~m chord. The wing is composed of three aluminum spars wrapped in carbon fiber to ensure rigidity. Fig. \ref{fig:WingSetup} shows the wing mounted to the ceiling of the tunnel test section. The root of the wing is connected to a 6-axis ATI IP15 SI165 load cell, measuring three component forces and moments. The measurement uncertainty at a $95$~\% confidence level as a percentage of the full-scale load is $1.25$~\%. The load cell protrudes from the inside of the test section, so there are no additional aerodynamic effects from any part of the system. Fig. \ref{fig:WingSetup} shows a flow splitter mounted within the test section $1$~mm from the wing tip to ensure the suppression of the wing tip vortex. An in-depth analysis of the forces and moments on the wing at different angles of attack is shown in \cite{thompson2023effects}.

\begin{figure*}[htbp]
    \centering
    \includegraphics[width=0.8\textwidth]{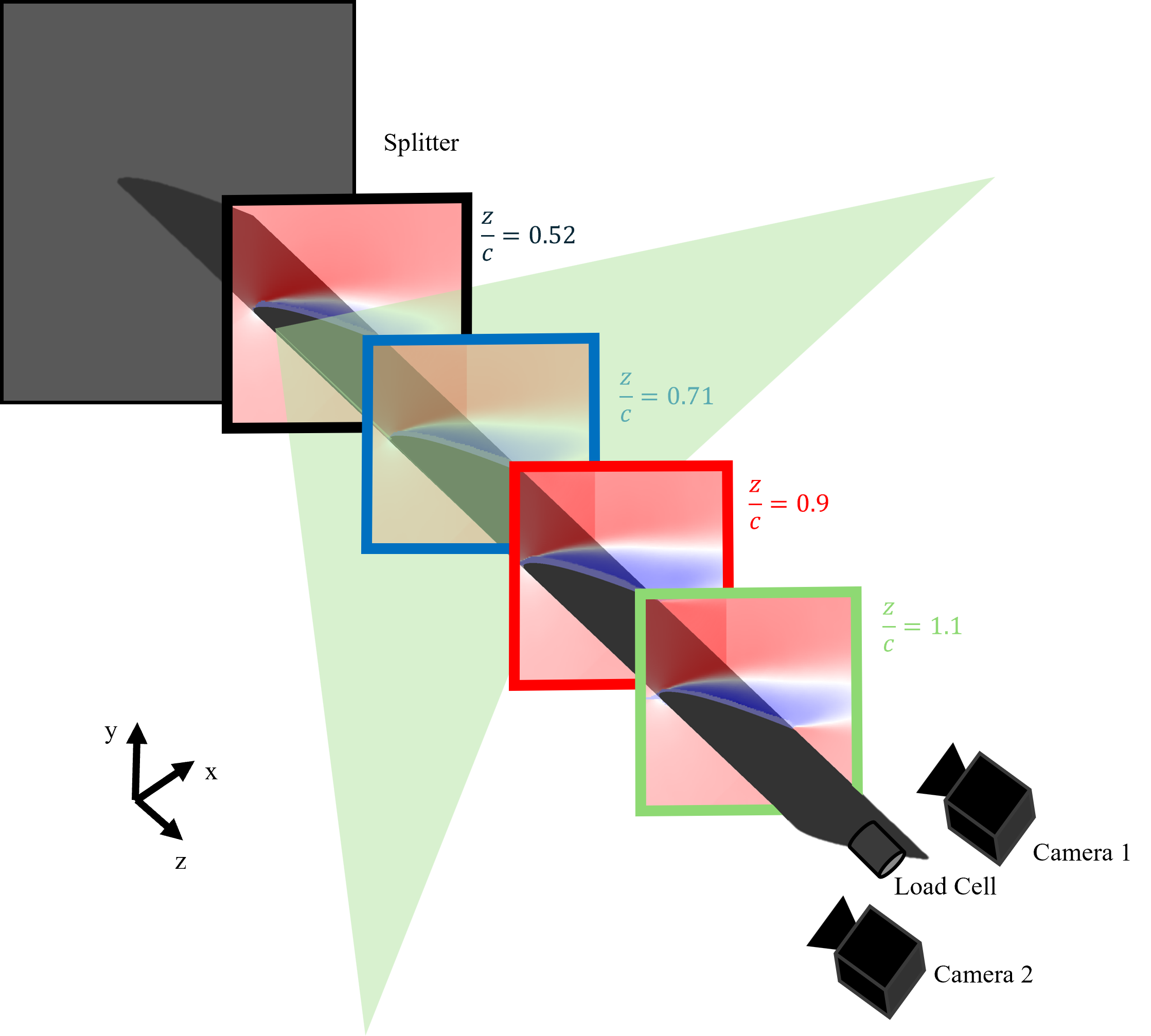}
    \caption{Schematic representation of the PIV setup.The green planes indicate the laser sheets, the black boxes the cameras, the grey rectangle the flow splitter, and the grey cylinder the 6-axis load cell. The colored boxes denote the specific spanwise locations of the experimental measurement planes at $z/c = 1.1$ (\protect\greenline), $z/c = 0.9$ (\protect\redline), $z/c = 0.71$ (\protect\blueline), and $z/c = 0.52$ (\protect\blackline), where the tip of the wing at the end plate is located at $z/c = 0$.}
    \label{fig:PIVSetup}
\end{figure*}

Fig. \ref{fig:PIVSetup} shows the planar PIV setup used to measure the flow field surrounding the wing. Two LaVision ImagerProLX cameras are mounted above the wing. Each camera is fitted with a $16$~MP CCD sensor producing a 3:2 aspect ratio and a pixel size of $7.4$~µm × $7.4$~µm, for a total size of $4872$~px × $3248$~px. The cameras have macro lenses with a focal length of $50$~mm. Two Nd:YAG Bernoulli lasers manufactured by Litron are mounted to a traverse on the side of the tunnel. The lasers are used to illuminate the wing's suction and pressure side in the chordwise plane at distances of $z/c = 1.1,~0.9,~0.71,$ and $0.52$. The lasers, shown in green in Fig. \ref{fig:PIVSetup}, ensure full coverage of the surrounding flow field. The lasers emit a double pulse of monochromatic light with a wavelength of $532$~nm and power of $200$~mJ shaped by mirrors and a cylindrical lens to form a light sheet $1$~mm thick.

The cameras and lasers are controlled by the PIV system developed by LaVision, which is implemented in the software \texttt{DaVis 8.4}. The software manages the data acquisition, storage, and timing of the triggers for the laser and cameras. A desktop computer with an inbuilt programmable timing unit manufactured by LaVision is used to ensure the relative timing of these signals. Following a sampling and convergence study, $1,000$ image pairs are captured at $0.6$~Hz. The 6-axis load cell is sampled at $10,000$~Hz for the duration of the PIV data acquisition.

As a result of two overlapping laser sheets paired with the shadows cast by the wing, there are regions of different light intensities within the camera's FOV. To mitigate the effects of the intensity mismatch, a method by \cite{mendez2017pod} is implemented where the first few POD modes within a threshold are subtracted from the raw images. The pre-processing method can be considered as background and noise subtraction as well as intensity correction.

Following the pre-processing procedure, a cross-correlation method is employed. An in-house \texttt{MATLAB} script is used for its superior processing speed. From a preliminary investigation, multiple passes with interrogation windows of size $128$~px $\times$ $128$~px, $64$~px $\times$ $64$~px and $32$~px $\times$ $32$~px are used, with an overlap of $75$~\%.

A third-order polynomial is used to correct for lens distortions and to map from the pixel coordinates to the physical coordinates. A natural-neighbor interpolation method is implemented such that the two object planes are projected onto a single structured grid \citep{sibson1981brief}. The size of the structured grid is determined from the average grid displacement of the two object planes, and the location of the structured grid is determined by reducing the mean distance from each point of the structured grid to each closest point of the object planes. The resulting FOV is approximately $1.5c \times 1.5c$.

\begin{table}[h]
  \centering
  \caption{Spatial resolution of the processed PIV vector fields at each spanwise measurement plane.}
  \renewcommand{\arraystretch}{1.4}
  \begin{tabular}{lcccc}
    \hline\hline
    Plane ($z/c$) & 1.1 & 0.9 & 0.71 & 0.52 \\
    \hline
    Resolution ($\times 10^{-3}c$) & 7.9 & 8.2 & 8.9 & 9.6 \\
    \hline\hline
  \end{tabular}
  \label{tab:piv_resolution}
\end{table}

Finally, a moving average is applied using a uniform $3 \times 3$ kernel. The resulting vector fields are then downsampled to demonstrate the robustness of the assimilation algorithm when sparse data are provided, retaining every third vector in both the streamwise and wall-normal directions, corresponding to an overall reduction in the spatial resolution by approximately $O(10)$. For variational approaches, \cite{thompson2024effect} demonstrate that the accuracy of the assimilation is insensitive to the resolution of the input data in $x$ and $y$. The spatial resolution of the processed vector fields at each spanwise plane is summarized in Table \ref{tab:piv_resolution}.

\subsection{Mean velocity fields}
\label{subsection:experimental_mean_velocity_fields}
\begin{figure*}[htbp]
    \centering
    
    % --- Top Figure (a): Schematic ---
    \begin{subfigure}{\textwidth}
        \centering
                \caption{}
        \includegraphics[width=0.85\textwidth]{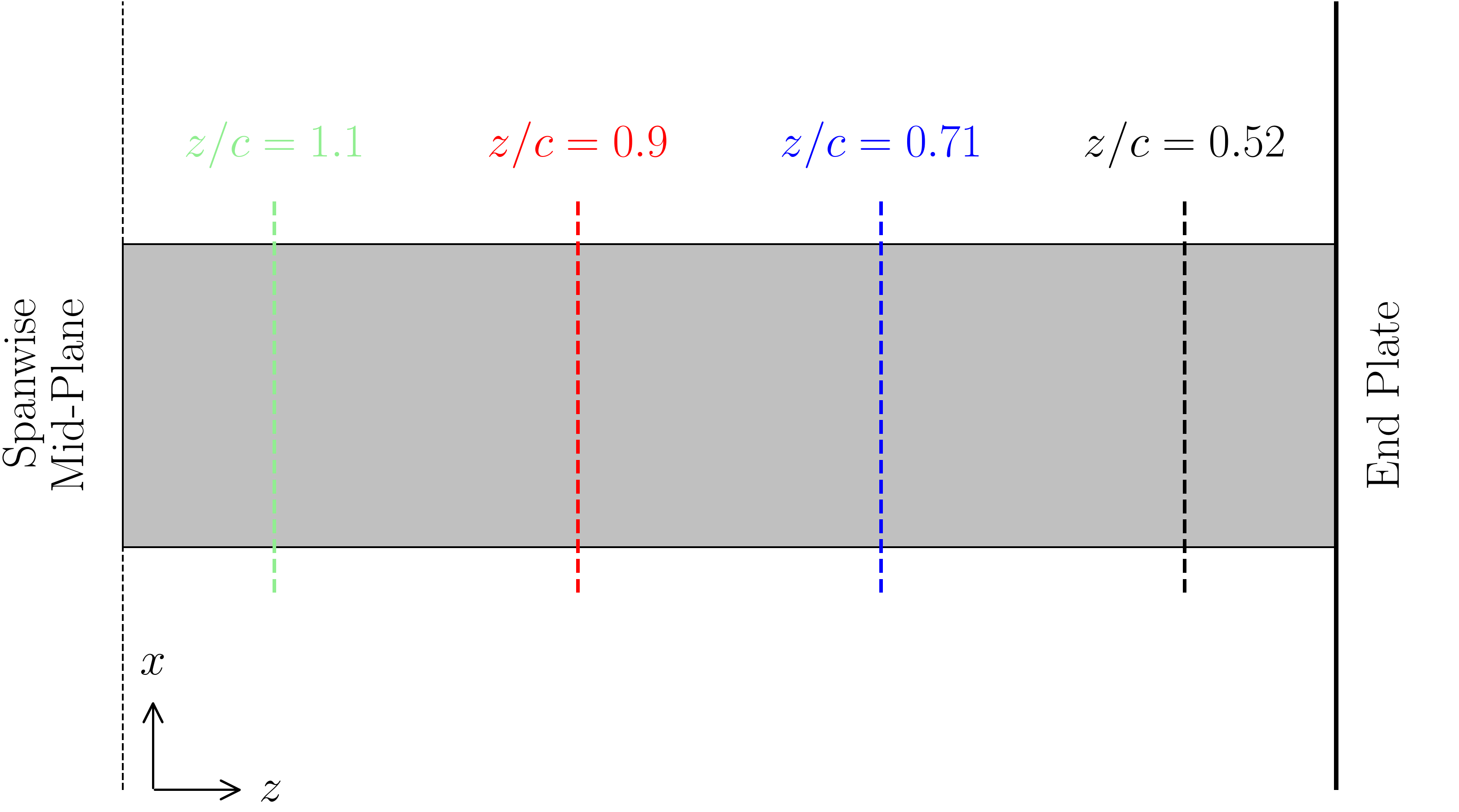}
        \label{fig:schematic}
    \end{subfigure}
    
    %%\vspace{1cm} % Adds breathing room between the two images
    
    % --- Bottom Figure (b): Experimental Data ---
    \begin{subfigure}{\textwidth}
        \centering
                \caption{}
        \includegraphics[width=0.85\textwidth]{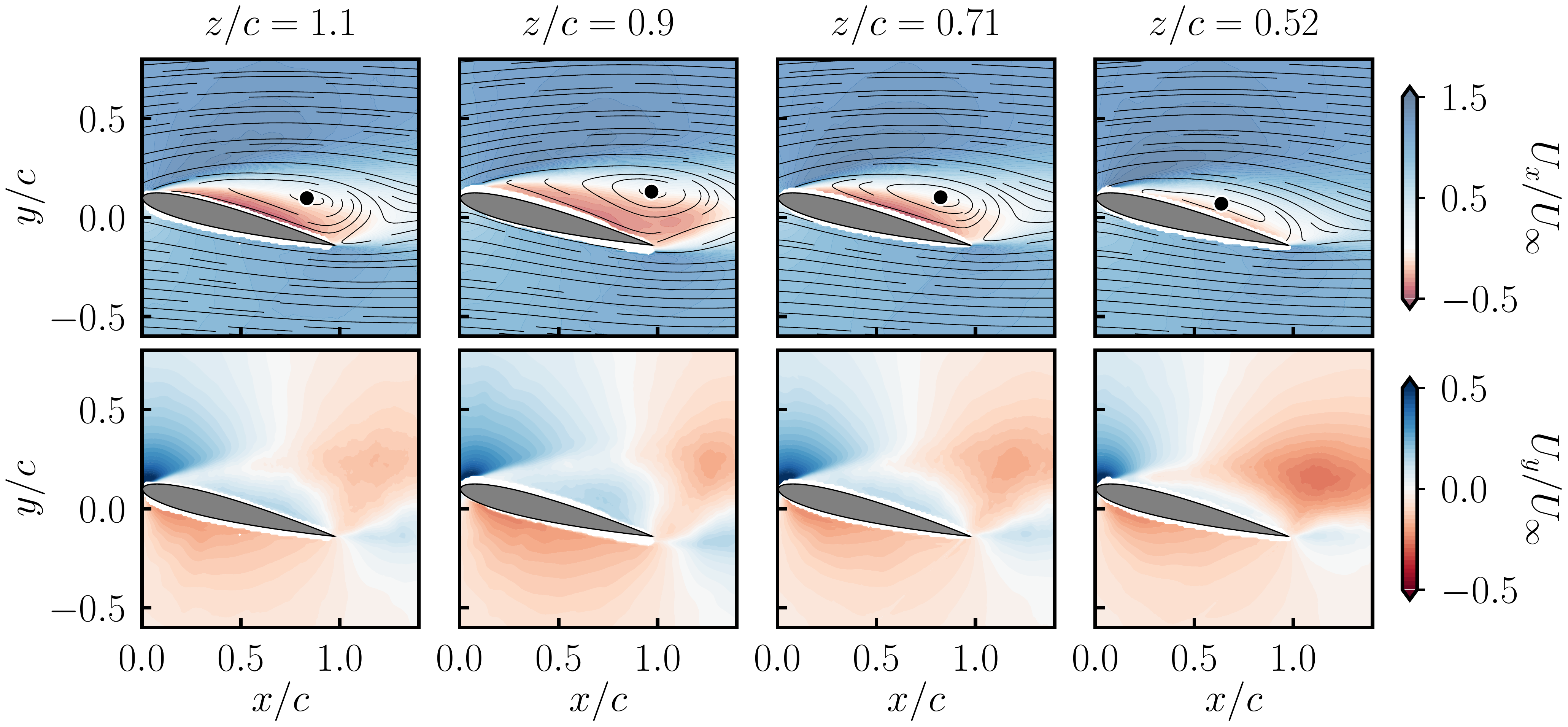}
        \label{fig:vel_components}
    \end{subfigure}
    
    \caption{\textit{(a)} Top-down schematic of the experimental setup illustrating the distribution of the spanwise measurement planes. \textit{(b)} Experimental time-averaged velocity fields at the four spanwise locations (from the end plate). The top row shows the normalized streamwise velocity ($U_x/U_\infty$) overlaid with streamlines. The core of the main recirculation bubble is shown with a filled black circle. The bottom row shows the normalized cross-stream velocity ($U_y/U_\infty$).}
    \label{fig:exp_setup_and_vel_components}
\end{figure*}

Fig. \ref{fig:schematic} shows a schematic of the wing semi-span on the $XZ$ plane, depicting the spanwise positions $z/c$ from the wingtip at which planar PIV data were acquired. The corresponding mean streamwise and wall-normal velocity fields are presented in Fig. \ref{fig:vel_components}, with the streamwise velocity field overlaid with streamlines and the core of the main recirculation bubble computed as the minimum of the mean velocity magnitude $|U| = \sqrt{U_x^2 + U_y^2}$, identified with a filled black circle. It is clear from Fig. \ref{fig:vel_components} that the mean velocity fields differ across the four spanwise locations. This is particularly evident in the extent of the recirculation bubble depicted by the streamlines. The spanwise plane along $z/c = 0.9$ shows the largest recirculation bubble denoting maximum separation, while the $z/c = 0.52$ plane has the smallest, showing the least separation. The core of the main recirculation bubble is furthest downstream for the $z/c = 0.9$ plane and closest to the leading edge on the $z/c = 0.52$ plane. \cite{neal2023three} showed that a modulation in the primary recirculation bubble core along the span is strong evidence for the presence of stall cells. The spanwise variation in recirculation bubble size and core location observed here serves as preliminary evidence for the presence of stall cells and is directly linked to the difference in separation points along the wingspan.

A closer examination of the velocity fields reveals a notable similarity between the mean velocity on planes $z/c = 1.1$ and $z/c = 0.71$. Both planes exhibit recirculation bubbles of comparable size, while the plane between them ($z/c = 0.9$) shows the largest separation. This pattern, two planes of moderate separation flanking a plane of maximum separation, is consistent with the spanwise structure reported in the stall cell literature \citep{winklemann1980, manolesos2014experimental, neal2023three}. The spanwise wavelengths reported for comparable configurations range from $\lambda_z \approx 1.25c$ \citep{sarras2024linear, liu2018numerical} to values predicted by the theoretical criterion of \cite{gross2015criterion} based on the local gradient of the lift polar. These observations, taken together with the spanwise modulation in recirculation bubble size and core location, provide strong evidence that the experimental flow field contains stall cells.

\subsection{Statistical stationarity in each measurement plane}
\label{subsection:statistical_stationarity}
\begin{figure*}[htbp]
    \centering
    \includegraphics[width=0.85\textwidth]{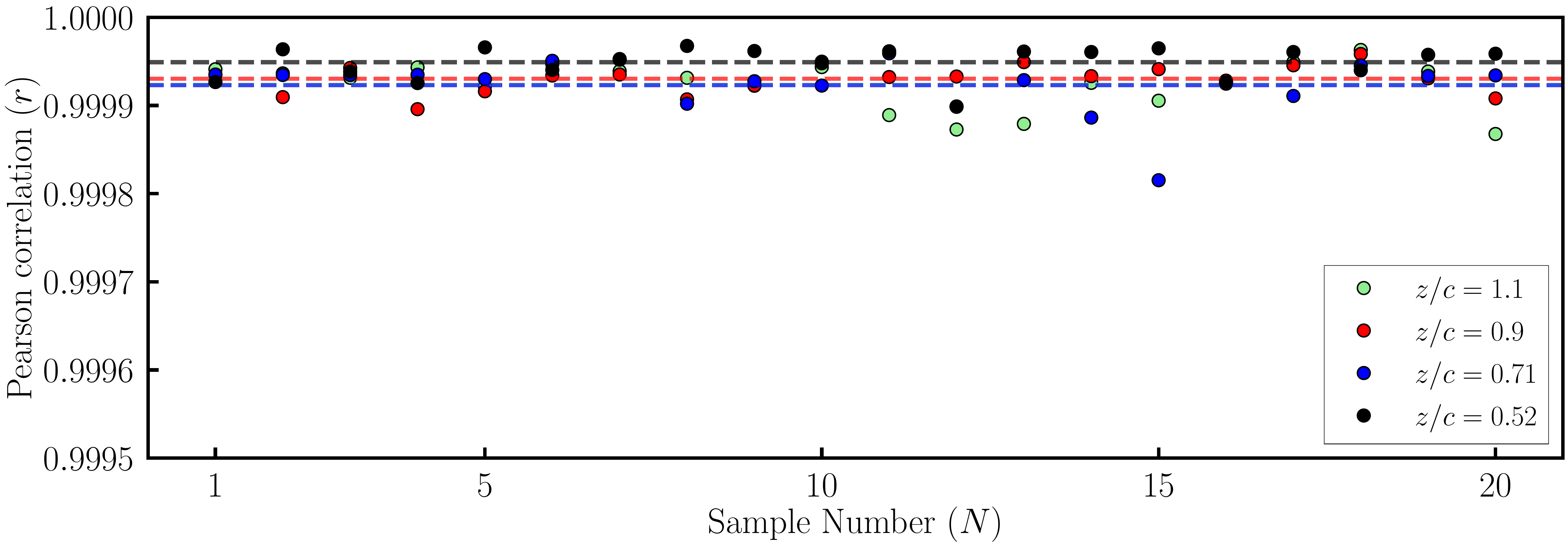}
    \caption{Statistical stationarity and structural stability of the spanwise velocity ensembles across the validation planes. Each discrete marker represents the Pearson correlation coefficient ($r$) between a random 500-snapshot subset (drawn with replacement) and the full 1000-snapshot ensemble mean. The dashed horizontal lines indicate the mean correlation level for each spanwise location. The exceptionally high correlation ($r > 0.9998$) and negligible variance between subsets confirm that the experimental stall cells are physically stationary. Marker colors correspond to the spanwise locations of experimental mean velocity fields (see Fig. \ref{fig:schematic} for the legend).}
    \label{fig:stationarity_convergence}
\end{figure*}

A natural question that arises is whether the stall cells observed in the present experiment are stationary or undergo spanwise movement over time. Several studies have reported that stall cells can be dynamic structures, exhibiting spanwise jostling, lateral migration, and intermittent formation and disappearance \citep{yon1998study, zarutskaya2005vortical, manolesos2014experimental, liu2018numerical}. Since the present PIV data are not time-resolved, the stationarity of the stall cells cannot be assessed through direct temporal tracking. Instead, a bootstrap procedure is employed. For each spanwise plane, $500$ snapshots are drawn at random with replacement from the full ensemble of $1000$ snapshots and the corresponding mean velocity field is computed. This process is repeated $20$ times. The similarity between each subset mean and the full-ensemble mean is quantified using the Pearson correlation coefficient, defined between two variables $A$ and $B$ as
\begin{align}
\label{equation:pearsons}
    \rho(A,B) = \frac{1}{N - 1} \sum_{i=1}^N \left( \frac{A_i - \mu_A}{\sigma_A}
    \right)\left(\frac{B_i - \mu_B}{\sigma_B}\right),
\end{align}
where $\mu$ and $\sigma$ denote the mean and standard deviation, respectively. The results are presented in Fig. \ref{fig:stationarity_convergence}. Across all four spanwise planes, the average correlation coefficient exceeds $r > 0.9998$ with negligible variance between subsets. This indicates that the mean velocity field is insensitive to which snapshots are included in the average, which would not be the case if the stall cells were translating or oscillating during the acquisition period. These results confirm that the stall cells in the present experiment are physically stationary.

A consideration that arises from the experimental procedure is that the four spanwise measurement planes were acquired in separate experimental runs, i.e. the planes are not measured simultaneously, as the optical setup required recalibration between spanwise locations. While the statistical stationarity analysis   confirmed that the stall cells are stationary within a single run, it cannot be guaranteed a priori that the stall cell occupies the same spanwise position across runs. However, several observations support the reproducibility of the stall cell position. First, the splitter plate imposes a fixed boundary condition at one end of the span, which serves to anchor the stall cell in a manner analogous to the localized spanwise disturbances used by \cite{manolesos2014experimental} to stabilize stall cells. Second, the flow conditions (Reynolds number, angle of attack, and freestream turbulence) are identical across all runs. Third, and most compellingly, the data from the four independently acquired planes exhibit a physically consistent spanwise structure: the planes at $z/c = 1.1$ and $z/c = 0.71$ show similar mean velocity fields, approximately symmetric about the most separated plane at $z/c = 0.9$. This pattern would be extremely unlikely to emerge if the stall cell position varied significantly between runs. The experimental data are therefore interpreted as representing the same statistically stationary stall cell structure captured across multiple runs.

\subsection{Spanwise coherence of flow dynamics}
\label{subsection:POD}
\begin{figure*}[htbp]
    \centering
    \includegraphics[width=0.65\textwidth]{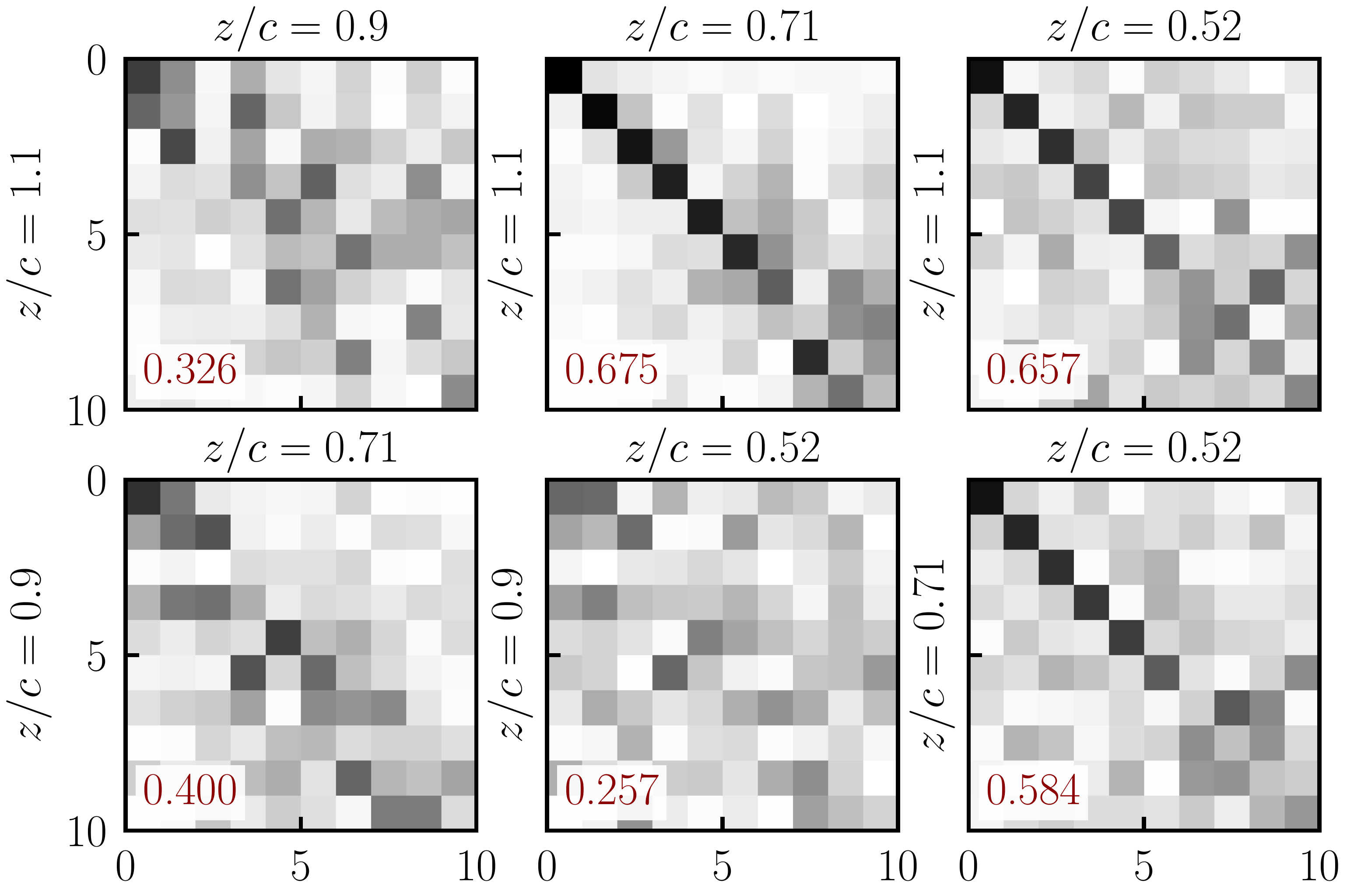}
    \caption{Pearson cross-correlation matrices of the first 10 spatial POD modes between pairs of spanwise measurement planes. The axes correspond to the mode indices ($0 \leq n, m \leq 10$) of the reference and target planes, with darker shades indicating a stronger correlation coefficient ($\rho$). The mean diagonal correlation ($\rho_{diag}$), annotated in the bottom-left corner of each panel, quantifies the overall spatial coherence and structural similarity of corresponding modes across the span.}
    \label{fig:pod_correlation}
\end{figure*}

Having established the differences in the mean velocity fields along the wingspan as preliminary evidence for the presence of stall cells, a closer examination of the flow dynamics is expected to further substantiate this observation. POD is a useful tool in fluid dynamics that decomposes the flow field into a linear combination of spatially orthogonal modes, ranked by their energy content, allowing the identification of dominant coherent structures \citep{lumley1967structure}. POD is applied to the instantaneous velocity snapshots obtained from the experiment on each of the four spanwise planes. The similarity between the dynamics of the POD modes across different spanwise planes is quantified using the Pearson correlation coefficient (Eq. \ref{equation:pearsons}). The steps followed to compute the POD modes and the Pearson correlation coefficients are outlined in Appendix \ref{appendix:pod}. The first $10$ POD modes are computed and ranked in order of modal energy for all four planes and the Pearson correlation coefficient is computed between every pair of planes for each of these $10$ modes. This results in a total of six correlation matrices (corresponding to the six unique pairings of the four planes), each of size $10 \times 10$, which are presented in Fig. \ref{fig:pod_correlation}.

The most striking observation is that the modes on the $z/c = 0.9$ plane correlate poorly with those on all other planes, indicating that the coherent structures at this spanwise location are distinctly different from the rest of the wingspan. In contrast, the modes on planes $z/c = 1.1$ and $z/c = 0.71$ are strongly correlated with each other, suggesting that these two locations experience similar fluctuation patterns. The plane at $z/c = 0.52$ exhibits moderate correlation with $z/c = 1.1$ and $z/c = 0.71$ but weak correlation with $z/c = 0.9$. These dynamic observations are consistent with the stall cell structure inferred from the mean flow analysis. The mean velocity data identified $z/c = 0.9$ as the location of maximum separation and showed that $z/c = 1.1$ and $z/c = 0.71$ have similar recirculation bubble characteristics. The POD analysis adds the complementary finding that the unsteady dynamics mirror this mean flow organization, with the stall cell center exhibiting distinct temporal behavior from the surrounding planes.

\section{Turbulence modeling and data assimilation}
\label{section:turbulence_data_assimilation}
Having described the experimental setup and characteristics of the mean velocity field in the previous section, the aim here is to introduce the RANS model and the DA methodology used to correct it. Section \ref{subsection:RANS} presents the baseline CFD model. Section \ref{subsection:data_assimilation} provides a detailed description of the variational formulation and proceeds to motivate the choice of the control variable, where a modification to the production term of the eddy viscosity is made. 
\subsection{Baseline CFD model}
\label{subsection:RANS}
Since the experimental data consist of time-averaged velocity fields and the DA operates on the mean flow, the governing equations are formulated in terms of the ensemble-averaged equations for an incompressible fluid,
\begin{gather}
\frac{\partial {U_i}}{\partial x_i} = 0, \\
\label{equation:RANS}
U_j \frac{\partial U_i}{\partial x_j} = -\frac{1}{\rho} \frac{\partial P^*}{\partial x_i} + \frac{\partial}{\partial x_j} \left( \nu \frac{\partial U_i}{\partial x_j} \right) - \frac{\partial  \tau_{ij}}{\partial x_j},
\end{gather}
where $U_i$ and $P^*$ are the mean velocity components and pressure, respectively, $\rho$ is the density of the fluid, $\nu$ is the kinematic viscosity, $\tau_{ij}$ is the Reynolds stress tensor, and $x_i$ represents the spatial coordinates. The Reynolds stress tensor $\tau_{ij} = \overline{u_i' u_j'}$ is defined as the averaged outer product of the fluctuating velocity components, presenting the well-known problem of closure. Here, the notation $( \cdot )'$ denotes fluctuations, representing the deviation of the instantaneous velocity from its ensemble average, expressed as
\begin{align}
u_i' = u_i - U_i,
\end{align}
where $u_i$ is the instantaneous velocity, and $U_i$ is the ensemble average. The overbar $\left(\,\overline{\,\cdot\,}\,\right)$ denotes Reynolds averaging. To model $\tau_{ij}$, the mean flow components are used within the Boussinesq hypothesis recasting Eq. \ref{equation:RANS} into
\begin{gather}
\label{equation:turb_model_rans}
U_j \frac{\partial U_i}{\partial x_j} + \frac{1}{\rho} \frac{\partial P^*}{\partial x_i} - \frac{\partial}{\partial x_j} \left( [\nu + \nu_t] \frac{\partial U_i}{\partial x_j} \right) = 0,
\end{gather}
where the eddy viscosity $\nu_t$ allows modeling of the Reynolds stress tensor through the Boussinesq approximation as $\tau_{ij} = -\nu_t \left(\partial U_i/\partial x_j + \partial U_j / \partial x_i \right) + \frac{2}{3}k\delta_{ij}$, where the isotropic term is absorbed into a modified pressure $P^*$. The eddy viscosity is computed by solving a transport equation. The \cite{spalart1992one} (SA) turbulence model (hereafter referred to as the baseline model) is used, which obtains $\nu_t$ from a surrogate variable $\tilde{\nu}$. 
\subsection{Data assimilation methodology}
\label{subsection:data_assimilation}
We expect that the solution to the RANS equations, as defined in Section \ref{subsection:RANS}, will deviate from experimental observations, necessitating the use of DA. To apply variational DA, we require the definition of a control variable and an objective function.  We use the field inversion framework devised by \cite{singh2017augmentation}. The production term of the SA turbulence model is augmented with a spatially-varying scalar field $\beta(x, y, z)$. The functional form of the augmented turbulence transport equation is given by:

\begin{align}
\label{equation:samodel}
    \frac{D \tilde{\nu}}{D t} = \beta (x,y,z) P(\tilde{\nu}, \mathbf{w}) + T(\nabla \tilde{\nu}, \mathbf{w}) - D(\tilde{\nu}, \mathbf{w}), 
\end{align}
where $\mathbf{w}$ is the vector of state variables such as mean velocity, pressure, and momentum flux, and $P$, $T$, and $D$ are the production, diffusive transport, and destruction terms, respectively. Bold-face symbols denote vectors throughout. It must be noted that setting $\beta(x,y,z) = 1$ throughout the domain recovers the baseline SA model without any modifications and is therefore the preferred choice for the initial condition of the optimizer. This results in the baseline SA model solution for the first optimization iteration, followed by incremental modifications. Alternatively, the optimizer can be initialized with a pre-computed baseline result. Either way, the availability of a baseline solution provides a good starting point for the optimizer, which is also recommended in \cite{franceschini2020mean}.

The objective function is the discrepancy between high-fidelity or experimental measurements, and RANS simulation mean velocity fields,
\begin{align}
\label{equation:mom_source_obj_func}
    J(\mathbf{u}, \beta) = \frac{1}{2} \|\mathcal{Q}(\mathbf{u}, \beta) - \tilde{\mathbf{Q}}\|_Q^2
\end{align}
where $\tilde{\mathbf{Q}}$ is a set of high-fidelity data/experimental measurements. Operator $\mathcal{Q} (\cdot)$ extracts the computational data in such a way that $\mathcal{Q}(\mathbf{u}) \in Q$ is a projection of the computational mean velocity onto the measurement space $Q$. \(\|\cdot\|_Q\) is the generic norm in the measurement space. It should be noted that while the objective function \(J\) is computed only within the region where high-fidelity data or experimental measurements are available, the control variable \(\beta\) is defined over the entire computational domain. Consequently, it possesses the same number of degrees of freedom as the mesh size.

Variational DA is now formulated as an optimization problem where the goal is to minimize an objective function subject to some constraints. This is mathematically written as
    \begin{eqnarray}
    \label{equation:constopti}
\min_{\beta \in \mathbb{R}^{n_{\beta}}} \quad & J(\beta),\\
    \label{equation:constopti1}
\textrm{s.t.} \quad & R(\mathbf{w}, \beta) = 0, \\
    \label{equation:constopti2}
\quad & \beta_L \leq \beta \leq \beta_U, 
\end{eqnarray}
where $n_{\beta}$ is the size of the control variable, $R$ is the governing equations that serve as constraints and $\beta_L$ and $\beta_U$ denote the lower and upper bounds, respectively, for the control variable.  The bounds $\beta_L$ and $\beta_U$ are determined through iterative tuning to balance optimizer flexibility and solution stability. Starting from tight bounds near zero, the range is gradually widened if convergence issues arise. Excessively narrow bounds can restrict the optimizer, while overly loose bounds may slow convergence or destabilize the primal solver by allowing unrealistically large control values. For our test case, $R$ represents the residual of the Navier--Stokes (NS) equations. In this formulation, the optimization is performed over the control variable $\beta$, while the state variable $\mathbf{w}$ is treated as an implicit function of $\beta$ and is obtained by solving the governing equations at each iteration. Equations \ref{equation:constopti} - \ref{equation:constopti2} form a non-linear, constrained minimization problem with equality and bound constraints and can be solved using gradient-based techniques. 

Gradient-based optimization techniques require the total derivative of the objective function with respect to the control variable (hereafter referred to as sensitivity). An efficient way to compute the sensitivity is by employing an adjoint method, which ensures that the computational cost remains independent of the number of control variables \citep{giannakoglou2008adjoint}. We use the discrete adjoint method in this study for computing the sensitivities. If $J$ and $R$ are a univariate representation of the objective function and residual of the NS equations, respectively, the sensitivity can be computed using
\begin{align}
\label{equation:discadjfinal}
   \frac{dJ}{d \beta} =\frac{\partial J}{\partial \beta} - \psi^T \frac{\partial R}{\partial\beta},
\end{align}
where $\psi^T$ is the transpose of the adjoint vector. The detailed derivation can be found in \cite{kenway2019effective}. The OpenFOAM library DAFoam is used to obtain the sensitivity \citep{he2018aerodynamic, he2020dafoam}. DAFoam's source code is enriched with AD-forward (ADF) and AD-reverse (ADR) implementations using CoDiPack \citep{SaAlGauTOMS2019}, enabling machine-precision gradient accuracy \citep{kenway2019effective}. The field inversion framework has been implemented and validated in DAFoam by \cite{bidar2022open} for canonical flow cases. 

Once the sensitivity is obtained, we use sequential quadratic programming (SQP) \citep{wilson1963simplicial} to perform the optimization. The SQP method solves an optimization problem by obtaining search directions from a sequence of quadratic programming (QP) sub-problems with linearized constraints. The sub-problem defines a quadratic model of a certain Lagrangian with linear equality and inequality constraints. A quasi-Newton method is used along with a line search filter that minimizes a penalty function to obtain search directions for the QP sub-problems. The initial set of iterates is then updated based on these search directions. We use SNOPT [Sparse Non-linear Optimizer, \citep{gill2005snopt}], which exploits the sparsity in the constraint Jacobian and employs a limited-memory BFGS (L-BFGS) approximation of the Hessian, allowing efficient handling of large-scale problems. The stopping criteria for the optimization are governed by the default convergence logic of the SNOPT optimizer, which monitors a combination of projected gradient norms and step sizes to determine when the optimization can no longer make meaningful progress.

We implement a nearest-neighbor operator $\mathcal{Q}$ similar to the one used in \cite{franceschini2020mean} to project the computational mean velocity data onto the experimental grid. The operator searches for the nearest computational grid point to a given experimental data point within the experimental field of view and samples the data at these points. The objective function can then be computed straightforwardly on these points. A more complex measurement operator that performs cell-volume averaging, taking into account cell-cell intersections, is implemented and used in \cite{thompson2024effect}, while a simpler operator that does not exploit cell-cell intersections is used in \cite{cadambi2026three}. The need to use such complex measurement operators is obviated by the result shown in \cite{franceschini2020mean} that a control variable optimized in the turbulence transport equation is robust to the sparsity of data. The added benefit of using the nearest-neighbor operator is that there is no need for a smoothing operator (which is used to transfer the discrepancy field onto the computational grid) since the objective function is being computed on the computational grid itself. 

\section{Case setup and baseline computation}
\label{section:case_setup_bsl_computation}
This section provides details on the computational case setup and results from the baseline SA model. Section \ref{subsection:case_setup} outlines the choice of the computational domain, discretization, solvers and boundary conditions. Section \ref{subsection:baseline} presents the mean velocity fields computed using the baseline model, along with a comparison with the experimental data.

\subsection{Case setup}
\label{subsection:case_setup}
The baseline computations are performed using the SA turbulence model, which is equivalent to setting $\beta(x,y,z) = 1$ throughout the domain, as described in Section \ref{subsection:data_assimilation}. The momentum and turbulence transport equations are discretized using a first-order upwind scheme. This choice ensures numerical stability and enables faster convergence of the adjoint solver. The matrix equations for pressure are solved using a geometric-algebraic multigrid (GAMG) method, while velocity and $\tilde{\nu}$ are solved using a smooth solver with a Gauss--Seidel smoother. The pressure-velocity coupling is handled through the semi-implicit method for pressure-linked equations (SIMPLE) algorithm using OpenFOAM's steady-state solver, \texttt{simpleFoam}. The residual tolerance is set to $10^{-6}$ for all flow variables. A tight tolerance on the residuals of the state variables is essential for ensuring efficient adjoint solver convergence. However, to allow for some flexibility in the solution process, a residual tolerance of up to $10^{-5}$ on any of the flow variables is accepted as a converged state. This has proven to be a good metric that allows a sufficiently converged primal solution while also ensuring that the adjoint solver converges.

\begin{figure*}[htbp]
    \centering
    \includegraphics[scale=0.25]{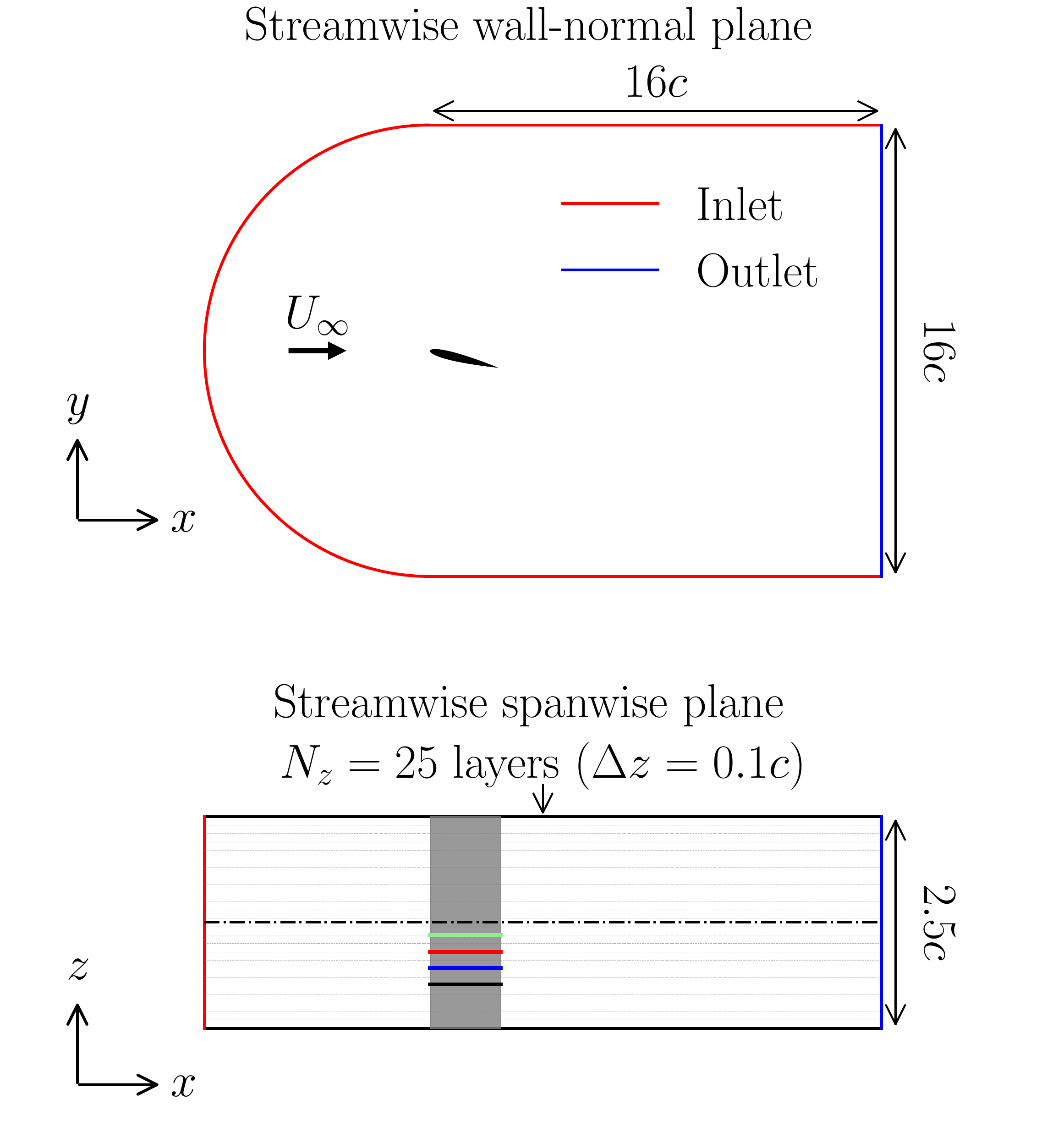}
    \caption{Schematic of the computational domain and boundary conditions. The top panel displays the streamwise wall-normal ($x$-$y$) plane, illustrating the C-grid topology with an upstream radius of $16c$ and a total vertical span of $16c$. The bottom panel shows the streamwise spanwise ($x$-$z$) plane, highlighting the 3D setup with a total spanwise domain of $2.5c$ discretized into $N_z = 25$ layers ($\Delta z = 0.1c$). The colored horizontal lines denote the specific spanwise locations of the experimental measurement planes (see Fig. \ref{fig:schematic} for the legend).}
    \label{fig:computational_domain_setup}
\end{figure*}

The computational grid used to perform the simulations is shown in Fig. \ref{fig:computational_domain_setup}. It is a C-type grid that extends $8c$ upstream, $16c$ downstream, and $8c$ above and below the leading edge of the airfoil. The majority of the domain is discretized using hexahedral elements, with a small region close to the airfoil surface using tetrahedral elements. The mesh is refined in the vicinity of the airfoil, and this refinement extends to the downstream boundary of the domain. The airfoil surface is discretized using $200$ points. No inflation layers are used to resolve the viscous sublayer, i.e. $y^+ < 1$ is not enforced, and wall functions are employed instead. This choice ensures that the problem remains computationally tractable, since resolving the viscous sublayer for a flow at $Re_c \approx 450{,}000$ is extremely demanding on computational resources. The mesh is generated in two steps. First, a 2D mesh consisting of a single layer of cells with spanwise extent $\Delta z = 0.1c$ is finalized. This is then extruded along the spanwise direction by $2.5c$ with $25$ cells along the span, resulting in an overall mesh size of approximately $1.5$ million cells. The spanwise cell spacing of $\Delta z = 0.1c$ is consistent with the resolution employed by \cite{liu2018numerical}, who successfully captured stall cells on a NACA 0012 at $Re_c = 10^6$ using the same spanwise discretization, and is finer than the coarsest resolution shown to sustain stall cells in \cite{manni2016numerical}. Moreover, since the present simulations are augmented with experimental data through DA, the corrected flow field is constrained by the observed physics, reducing the sensitivity of the results to the spanwise mesh resolution alone \citep{thompson2024effect}.

\begin{table}[htbp]
  \centering 
  \caption{\label{tab:bc_bsl} Boundary conditions for baseline SA model.}
  \renewcommand{\arraystretch}{2}
  \begin{tabular}{lcccc}
    \hline\hline
    & Inlet & Outlet & Front \& Back & Airfoil \\
    \hline
    $U$ & $U_{\infty}$ & $\dfrac{\partial U}{\partial n} = 0$ 
        & $\hat{n}\cdot\vec{U}=0$ \, \& \, $\dfrac{\partial U_\parallel}{\partial n} = 0$ 
        & no slip \\[6pt]
    \hline
    $p$ & $\dfrac{\partial p}{\partial n} = 0$ & $0$ 
        & $\dfrac{\partial p}{\partial n} = 0$ 
        & $\dfrac{\partial p}{\partial n} = 0$ \\[6pt]
    \hline
    $\tilde{\nu}$ & $\tilde{\nu}_{\infty}$ 
        & $\dfrac{\partial \tilde{\nu}}{\partial n} = 0$ 
        & $\dfrac{\partial \tilde{\nu}}{\partial n} = 0$ 
        & $\tilde{\nu} = 0$ \\[6pt]
    \hline
    $\nu_t$ & \texttt{calculated} & \texttt{calculated} 
        & $\dfrac{\partial \nu_t}{\partial n} = 0$ 
        & \texttt{nutUSpaldingWallFunction} \\
    \hline\hline
  \end{tabular}
\end{table}

The boundary conditions applied to velocity, pressure, $\nu_t$, and $\tilde{\nu}$ are listed in Table \ref{tab:bc_bsl}. A velocity-inlet, pressure-outlet boundary condition is prescribed, setting fixed values for velocity and pressure at the inlet and outlet, respectively, following the standard approach for incompressible flows. At the inlet, $\tilde{\nu}_{\infty}$ is prescribed in the range $3\nu_{\infty}$ to $5\nu_{\infty}$ as recommended by \cite{spalart1992one}, with a zero-gradient condition imposed at the outlet. Since the transport equations are solved for $\tilde{\nu}$ as outlined in Eq. \ref{equation:samodel}, the eddy viscosity $\nu_t$ at the inlet and outlet is designated as \texttt{calculated}, meaning that it is derived from the local value of $\tilde{\nu}$ using the relation $\nu_t = \tilde{\nu} f_{v1}$ provided in \cite{spalart1992one} rather than being independently prescribed. Since the viscous sublayer is not resolved, a wall function approach is adopted at the airfoil surface. Specifically, the \texttt{nutUSpaldingWallFunction} is applied to $\nu_t$ at the wall, which is based on the continuous law of the wall formulation of \cite{spalding1961single} and provides a smooth blending between the viscous sublayer and the log-law region without requiring an explicit switch. The modified eddy viscosity is set to $\tilde{\nu} = 0$ at the wall, and $\nu_t$ is computed from $\tilde{\nu}$ at interior cells using the same relation, while at the wall its value is overridden by the wall function. The lateral faces of the domain are treated as slip boundaries with zero normal velocity and zero normal gradients of the tangential velocity, pressure, and turbulence quantities. This condition better represents the end conditions imposed by the splitter plate in the experiment and provides a closer approximation to the physical setup than periodic boundary conditions.

\subsection{Baseline mean velocity field}
\label{subsection:baseline}
The results of the mean velocity field computed using the baseline SA model are presented here. The $L_1$ norm is used as a metric to quantify the misfit between the baseline computation and the experiment. Since there is a mismatch between the computational and experimental grids, the computational data are linearly interpolated onto the downsampled experimental grid. The closest spanwise plane from the computational grid corresponding to each spanwise location where experimental data are available is first identified to within a small tolerance. The $L_1$ norm is then computed using the relation
\begin{align}
\label{eqn:L1NormPaper}
L_1 = \frac{|\mathcal{Q}(U_x) - \hat{U}_x| + |\mathcal{Q}(U_y) - \hat{U}_y|}{U_{\infty}},
\end{align}
where $\hat{U}_x$ and $\hat{U}_y$ are the mean streamwise and wall-normal velocities, respectively, from the experiment and $\mathcal{Q}$ is the projection operator, in this case a linear interpolator. The $L_1$ norm is chosen as it provides a robust measure of the misfit that is less sensitive to localized outliers in the velocity field. All subsequent $L_1$ norm fields presented herein use the relation defined in Eq.~\ref{eqn:L1NormPaper}.

\begin{figure*}[htbp]
    \centering
    \includegraphics[scale=0.17]{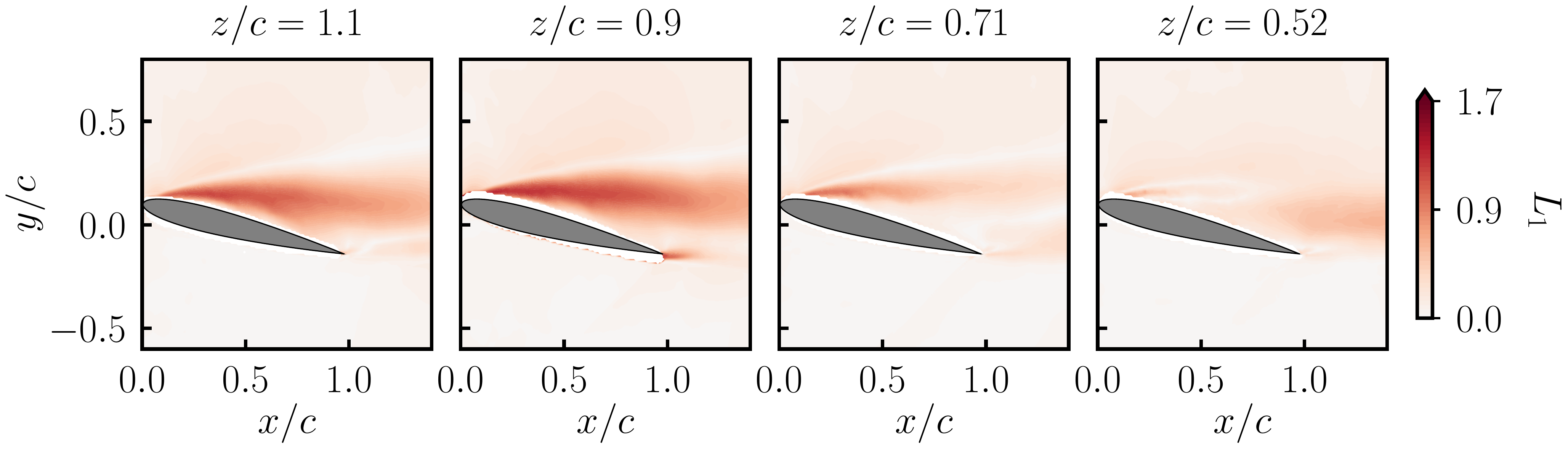}
    \caption{Spatial distribution of the $L_1$ error norm between the baseline CFD predictions and the experimental mean velocity fields at four spanwise locations. The $L_1$ norm is calculated as  $|U_{x, \text{CFD}} - U_{x, \text{Exp}}| + |U_{y, \text{CFD}} -  U_{y, \text{Exp}}|$, where the velocities are normalized by the  free-stream velocity $U_\infty$.}
    \label{fig:baseline_l1_norm}
\end{figure*}

The spatial distribution of the $L_1$ norm computed on all four planes of experimental data is presented in Fig. \ref{fig:baseline_l1_norm}. The $L_1$ norm is highest on planes $z/c = 1.1$ and $0.9$ and progressively decreases on planes closer to the splitter plate. More specifically, planes $z/c = 1.1$ and $0.9$ exhibit large errors that extend from the leading edge shear layer into the airfoil wake. A shear layer error persists on the $z/c = 0.71$ plane, while the plane closest to the splitter plate ($z/c = 0.52$) shows minimal error, confined to the trailing edge region. While it is expected that RANS turbulence models fail to predict the extent of recirculation in the wake of a stalled airfoil, it is notable that this error exhibits a clear spanwise distribution. One possible interpretation is that the baseline model performs adequately on planes where the experimental data indicates relatively mild separation (such as $z/c = 0.52$, see Fig. \ref{fig:exp_setup_and_vel_components}), but fails on planes where the separation is most severe. If this were the case, the spanwise variation in the $L_1$ norm would simply reflect the expected degradation of RANS accuracy with increasing separation. However, a closer examination of the baseline flow field is needed to determine whether this explanation is sufficient.

\begin{figure*}[htbp]
    \centering
    \includegraphics[scale=0.17]{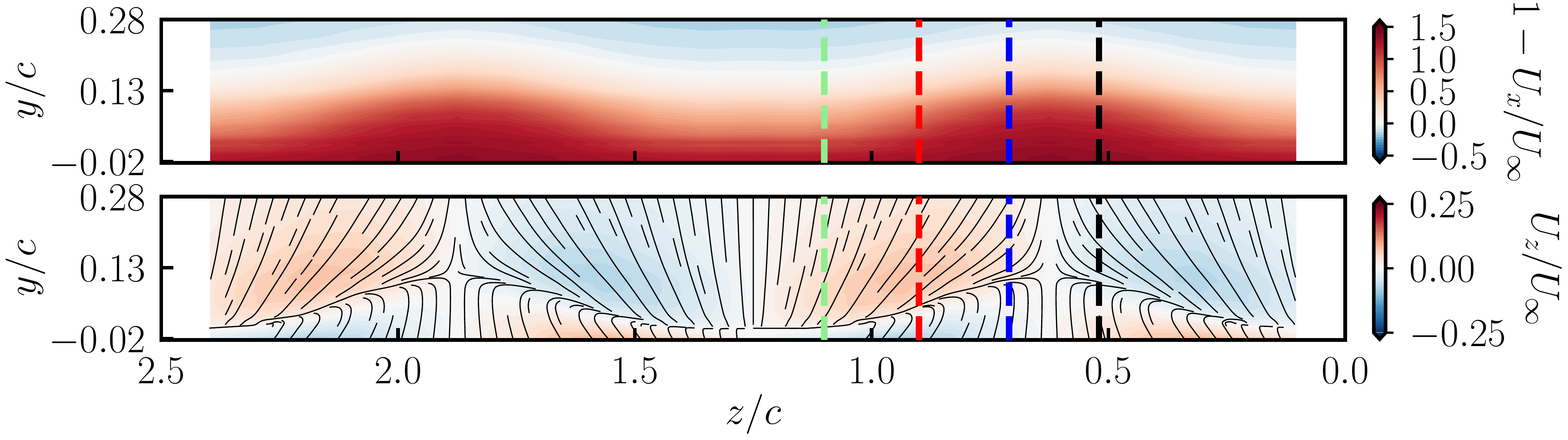}
    \caption{Baseline suction-side flow characteristics at $x/c = 0.7$.  The top panel presents the mean streamwise velocity deficit ($1 - U_x/U_\infty$), highlighting the spatial extent of the stall cells. The bottom panel displays the spanwise velocity ($U_z/U_\infty$) overlaid with in-plane secondary flow streamlines ($U_y, U_z$), illustrating the three-dimensional focal structures driving the separation. The dashed vertical lines indicate the position of the four spanwise experimental measurement planes used for reference (see Fig. \ref{fig:schematic} for plane legend).}
    \label{fig:suction_flow_baseline}
\end{figure*}

To this end, the wake deficit $1 - U_x/U_\infty$ at $70~\%$ of the chord from the leading edge ($x/c = 0.7$) is presented in Fig. \ref{fig:suction_flow_baseline}, where larger positive values indicate stronger recirculation, along with the spanwise velocity superimposed with in-plane ($YZ$) streamlines. The spanwise locations where experimental data are available are shown using dashed lines, with colors matching those in Fig.~\ref{fig:exp_setup_and_vel_components}. Two regions of large wake deficit, corresponding to large recirculation bubbles, are visible, sandwiching a region of lower wake deficit that indicates a smaller recirculation bubble and presumably more delayed separation. The in-plane streamlines and spanwise velocity contours show weak three-dimensional recirculation on this plane. Interestingly, the largest wake deficit does not coincide with the spanwise planes $z/c = 1.1$ and $0.9$ where the experimental data shows the most separation. This suggests that the large $L_1$ errors on these planes are not solely due to an under-prediction of the separation bubble, but also reflect a spanwise displacement of the region of strongest recirculation in the baseline relative to the experiment.

\begin{figure*}[htbp]
    \centering
    \includegraphics[scale=0.17]{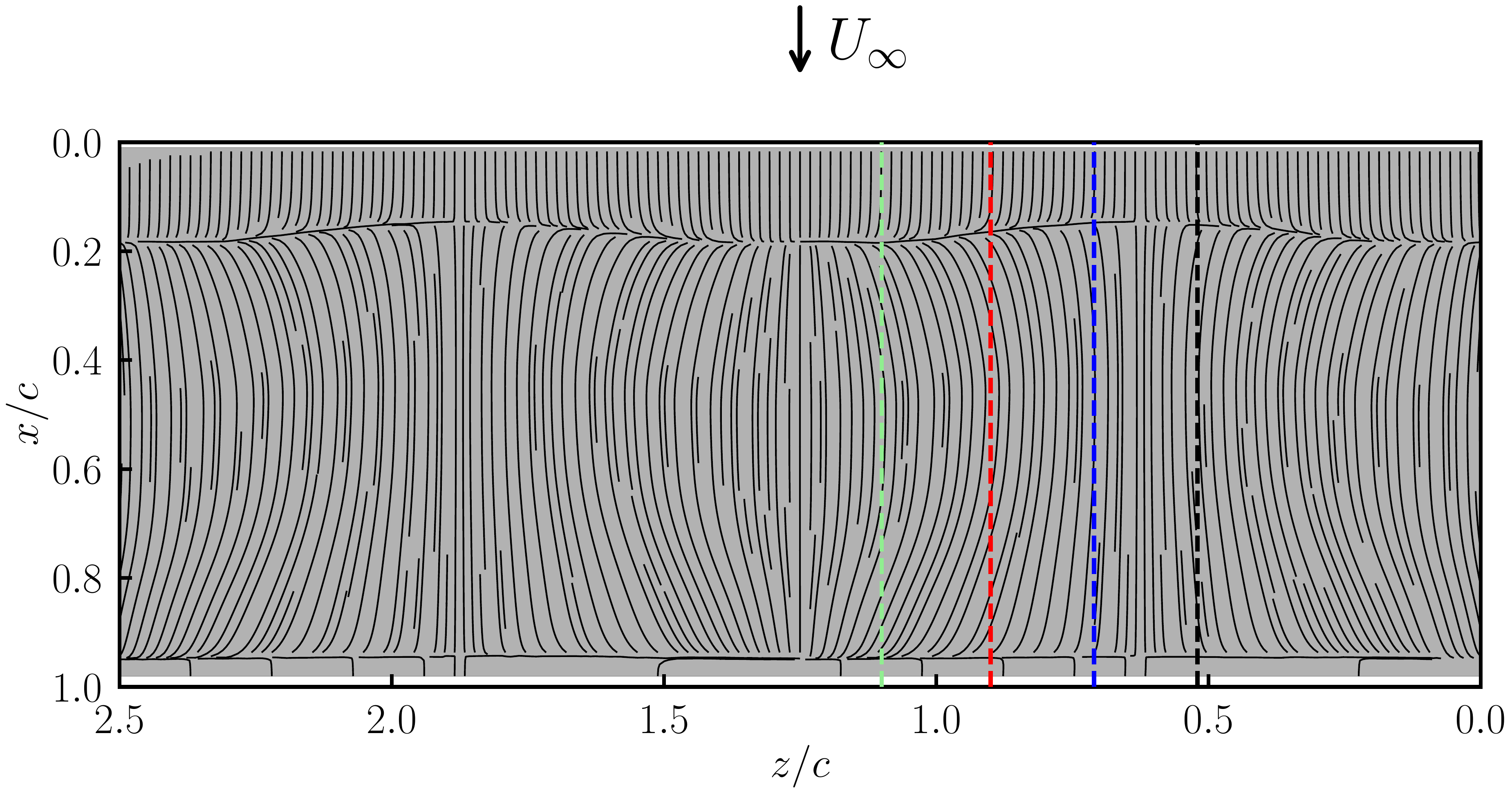}
   \caption{Streamlines of streamwise ($\tau_x$) and spanwise ($\tau_z$) components of wall shear stress obtained from baseline simulation using RANS SA model illustrating the surface flow pattern. Dashed lines indicate the experimental spanwise measurement planes (see Fig.~\ref{fig:schematic} for plane legend).}
    \label{fig:baseline_surface_streamlines}
\end{figure*}

The surface topology of the baseline computation provides a definitive assessment of whether these features constitute stall cells. Fig. \ref{fig:baseline_surface_streamlines} presents surface streamlines on the suction side of the airfoil, computed from the streamwise and spanwise components of the wall shear stress ($\tau_x$, $\tau_z$). This visualization is analogous to the oil-flow technique commonly employed in experimental studies of stall cells \citep{moss1971two, winkleman1982, manolesos2014experimental}. While there is some evidence of three-dimensionality in the form of a slight modulation in the separation line, the surface topology does not exhibit the defining features of stall cells. In particular, the counter-rotating vortices around focal points on the surface, that are characteristic of stall cells, are absent, and the separation line remains largely uniform along the span. The baseline SA model therefore fails to reproduce the stall cell topology observed in the experiment, despite the weak three-dimensionality visible in the cross-plane data of Fig. \ref{fig:suction_flow_baseline}. This establishes the need to inform the baseline SA model with experimental data in order to reconstruct a flow field consistent with the presence of stall cells.

\section{Stall-cell reconstruction}
\label{section:stall_cell_reconstruction}
This section presents a systematic study of the stall cell reconstruction using limited experimental data. It remains unclear how much experimental data is sufficient to recover stall cell structures in the corrected flow field. To address this, single-plane and dual-plane assimilation cases are investigated. Section \ref{subsection:data_assimilated_mean_fields} presents the convergence of the optimizer and the performance of each assimilation case, assessed using the $L_1$ norm and the Pearson correlation coefficient. The spatial distribution of the $L_1$ norm improvement for the selected cases is examined in Section \ref{subsection:spatial_l1_distribution}. Their suction-side flow structure and surface topology are then analyzed in Section \ref{subsection:wake_deficit_surface_topology}.

\subsection{Convergence and assimilation performance}
\label{subsection:data_assimilated_mean_fields}

Six assimilation cases are considered: four single-plane cases, each using one of the four available experimental planes as reference data, and two dual-plane cases, each using two planes simultaneously. The cases are labeled S1--S4 for the single-plane cases and D1--D2 for the dual-plane cases, in order of decreasing spanwise distance from the wingtip: S1 ($z/c = 1.1$), S2 ($z/c = 0.9$), S3 ($z/c = 0.71$), S4 ($z/c = 0.52$), D1 ($z/c = 1.1$ and $0.52$), and D2 ($z/c = 1.1$ and $0.9$). This nomenclature, summarized in Table \ref{tab:case_nomenclature}, is used throughout the remainder of the paper. The dual-plane cases were chosen to test two limiting scenarios: D1 uses the two planes that are furthest apart in spanwise distance, while D2 uses two planes that are close together but exhibit the largest difference in mean separation extent. In both cases, the mean velocity fields on the selected planes are sufficiently dissimilar to provide complementary information to perform DA. Other combinations were excluded either because their mean velocity fields are too similar (e.g.\ S1 and S3) or because they would be redundant with an existing case (e.g.\ S2 and S3). For the single-plane cases, the objective function is evaluated on the corresponding reference plane. For the dual-plane cases, the objective function is the sum of the misfits computed on both reference planes, normalized together by the corresponding baseline value.

\begin{table}[h]
  \centering
  \caption{Summary of assimilation cases.}
  \renewcommand{\arraystretch}{1.4}
  \begin{tabular}{lcc}
    \hline\hline
    Case & Type & Data plane(s) ($z/c$) \\
    \hline
    S1 & Single-plane & 1.1 \\
    S2 & Single-plane & 0.9 \\
    S3 & Single-plane & 0.71 \\
    S4 & Single-plane & 0.52 \\
    D1 & Dual-plane & 1.1, 0.52 \\
    D2 & Dual-plane & 1.1, 0.9 \\
    \hline\hline
  \end{tabular}
  \label{tab:case_nomenclature}
\end{table}

%\begin{figure}[!h]
%    \centering
%    \includegraphics[scale=0.17]{figures/figure_10.pdf}
%    \caption{Convergence history of the normalized objective function ($J/J_0$) with respect to the major iterations ($N$). Single-plane cases are shown as S1 (\protect\greenline), S2 (\protect\redline), S3 (\protect\blueline), and S4 (\protect\blackline). The dual-plane cases are shown as D1 (\protect\brownline) and D2 (\protect\cyanline).}
%    \label{fig:objective_convergence}
%\end{figure}

\begin{figure*}[htbp]
    \centering
    \includegraphics[scale=0.17]{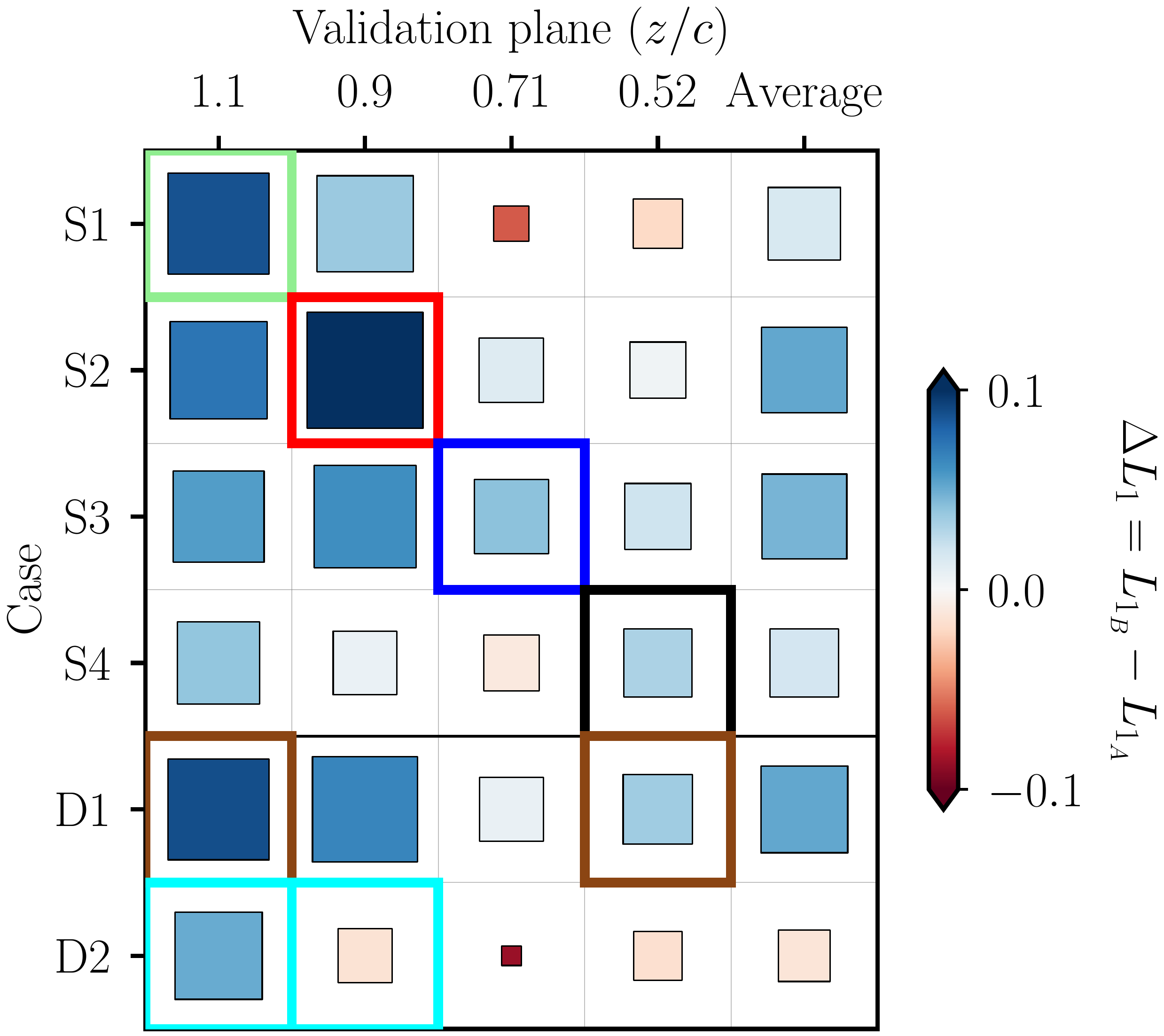}
\caption{Waterfall plot of assimilation performance across the four spanwise validation planes. The rows correspond to cases S1--S4 (single-plane) and D1--D2 (dual-plane) as defined in Table \ref{tab:case_nomenclature}. Each column corresponds to a validation plane ($z/c = 1.1$, $0.9$, $0.71$, $0.52$ from left to right), with the final column showing the row-averaged performance. The color of each tile represents $\Delta L_1 = L_{1_B} - L_{1_A}$ (Eq. \ref{eqn:waterfall_l1}), where darker blue denotes greater improvement. The area of each tile is scaled by $\Delta r = r_A - r_B$ (Eq. \ref{eqn:waterfall_pearsons}), where larger tiles denote greater improvement in structural similarity. Colored borders indicate the reference data planes used for assimilation in each case.}
    \label{fig:waterfall_absolute}
\end{figure*}

All six assimilation cases converge to a reduced objective function value, confirming that the optimizer successfully reduces the misfit between the assimilated and experimental velocity fields on the reference plane(s). However, the objective function is evaluated only on the reference data planes and therefore does not capture the accuracy of the reconstruction on off-data planes, nor does it convey the improvement in the spatial structure of the flow. To assess the reconstruction quality across all four spanwise planes on a common basis, two complementary metrics are combined into the waterfall plot shown in Fig. \ref{fig:waterfall_absolute}. The plot consists of six rows corresponding to the six assimilation cases and five columns corresponding to the four spanwise planes and a column-averaged value. Each tile encodes two quantities simultaneously.

The first quantity is the difference in the $L_1$ norm between the 
baseline and the assimilated fields,
\begin{align}
\label{eqn:waterfall_l1}
    \Delta L_1 = L_{1_B} - L_{1_A},
\end{align}
where $L_{1_B}$ and $L_{1_A}$ are the $L_1$ norms (Eq. \ref{eqn:L1NormPaper}) of the baseline and assimilated fields with respect to the experiment, respectively. A positive value of $\Delta L_1$ indicates that the assimilated field is closer to the experiment than the baseline, and is represented by a darker blue shade in Fig. \ref{fig:waterfall_absolute}.

The second quantity is the improvement in the Pearson correlation coefficient of the reconstructed mean velocity field, calculated as
\begin{align}
\label{eqn:waterfall_pearsons}
    \Delta r = r_A - r_B,
\end{align}
where $r_A$ and $r_B$ are the Pearson correlation coefficients (Eq. \ref{equation:pearsons}) between the experiment and the assimilated and baseline fields, respectively. A positive value of $\Delta r$ indicates improved structural similarity with the experiment. This quantity is mapped to the area of each tile: a larger tile denotes a greater improvement in correlation. The waterfall plot is read by selecting a row (case) and examining the two metrics across the four planes and their average in the final column. The tiles with colored borders indicate the reference data planes used for assimilation in each case. The metrics on these bordered tiles correspond directly to the objective function, since they quantify the misfit on the same planes that the optimizer was driven to minimize.

The waterfall plot in Fig. \ref{fig:waterfall_absolute} is now examined row by row, from case S1 (topmost) to case D2 (bottommost), to assess the performance of each assimilation case on the basis of the mean velocity reconstruction on off-data planes using the two metrics defined earlier: the $L_1$ norm difference $\Delta L_1$ and the Pearson correlation coefficient improvement $\Delta r$. The final column in each row represents the average performance across all four validation planes, which includes both the reference data plane(s) (highlighted by colored borders) and the off-data planes.

Case S1 shows a strong improvement on the reference plane and the adjacent plane ($z/c = 0.9$), but does not improve the mean velocity field on the two planes closer to the splitter plate ($z/c = 0.71$ and $z/c = 0.52$). In fact, providing data along $z/c = 1.1$ results in a slight worsening of both the $L_1$ norm and the correlation on these two planes. However, since the improvement on the first two planes is substantial, the row-averaged performance remains positive. Case S2 shows a similar pattern, with a strong improvement on the reference plane and the adjacent plane, but with the notable difference that the two planes closer to the splitter plate are not adversely affected.

Cases S3 and S4 exhibit different behavior. Despite the reference data being provided on planes where the baseline misfit is already small (see Fig. \ref{fig:baseline_l1_norm}), both cases produce a noticeable improvement on the two planes furthest from the splitter plate ($z/c = 1.1$ and $z/c = 0.9$). The improvement on the reference planes themselves is modest, which is expected given the small baseline misfit. These two planes appear resistant to change, yet the data provided on them enforces a measurable improvement on the off-data planes further from the boundary. Both S3 and S4 are therefore selected for further analysis in Section \ref{subsection:spatial_l1_distribution} to examine how the spanwise location of the reference data affects the reconstructed stall cell topology.

For the dual-plane cases, D1 provides data on the two planes that are furthest apart ($z/c = 1.1$ and $z/c = 0.52$). The improvement is concentrated on the two reference planes, with the off-data planes ($z/c = 0.9$ and $z/c = 0.71$) showing only minor changes. Case D2, which provides data on $z/c = 1.1$ and $z/c = 0.9$, shows a strong improvement on both reference planes. The off-data planes ($z/c = 0.71$ and $z/c = 0.52$) show only minor changes, consistent with the trend observed in the single-plane cases. D2 performs slightly better than D1 in terms of the row-averaged metrics and is therefore also selected for further analysis in Section \ref{subsection:spatial_l1_distribution} alongside S3 and S4.

\subsection{Spatial distribution of assimilation improvement}
\label{subsection:spatial_l1_distribution}
\begin{figure*}[htbp]
    \centering
    
  \begin{subfigure}{\textwidth}
        \centering
        \caption{}
        \includegraphics[scale=0.17]{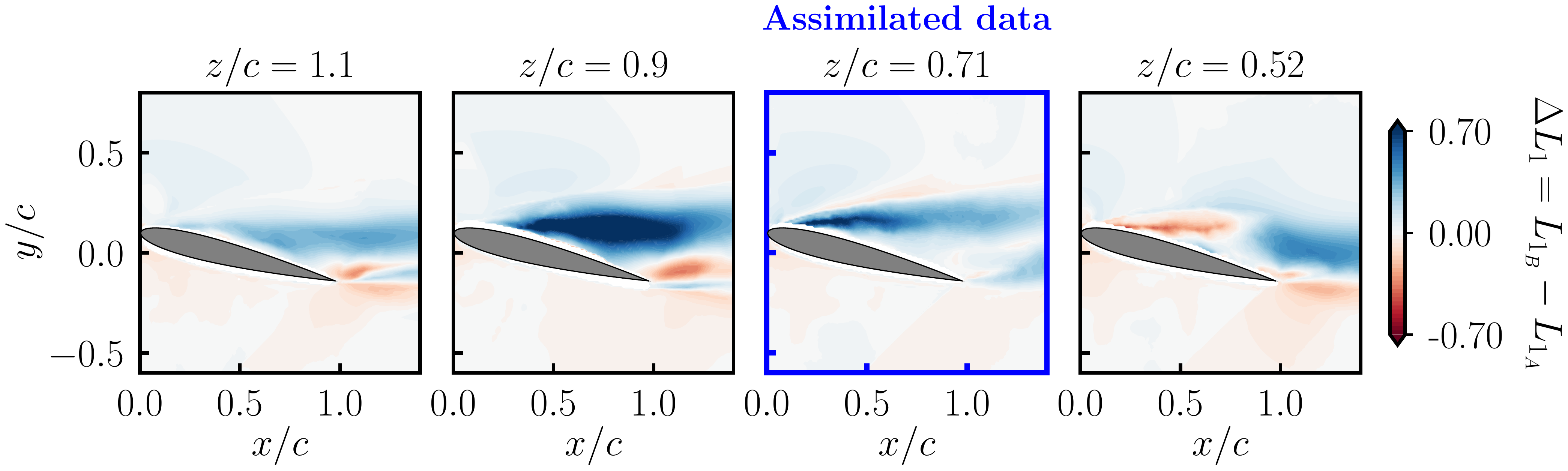}
        \label{fig:spatial_l1_single_0p7}
    \end{subfigure}
    
    \vspace{-0.5cm}
        \begin{subfigure}{\textwidth}
        \centering
        \caption{}
        \includegraphics[scale=0.17]{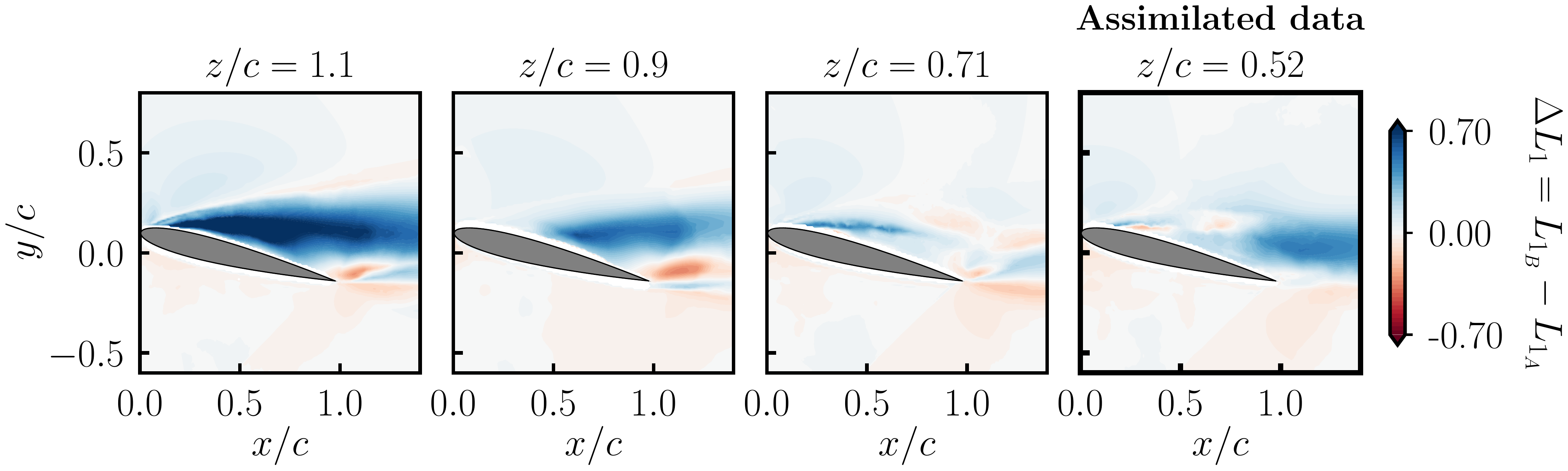}
        \label{fig:spatial_l1_single_0p52}
    \end{subfigure}

    \caption{Spatial distribution of $\Delta L_1 = L_{1_B} - L_{1_A}$ across the four validation planes for the single-plane cases (a) S3 and (b) S4 (see Fig. \ref{fig:waterfall_absolute} for nomenclature). Blue regions indicate improvement over the baseline and red regions indicate localized degradation. Colored bounding boxes denote the reference data planes (see Fig. \ref{fig:schematic} for the legend).}
    \label{fig:spatial_l1_improvement_single}
\end{figure*}

To understand which regions of the flow have improved and which have not, the spatial distribution of $\Delta L_1$ for the single-plane cases is presented in Fig. \ref{fig:spatial_l1_improvement_single}, with the reference data plane highlighted using the same color convention as before. This is the same metric that is presented as an integrated quantity in the waterfall plot of Fig. \ref{fig:waterfall_absolute}. Both single-plane cases show improvement over the baseline along planes $z/c = 1.1$ and $z/c = 0.9$, even though data are not provided on these planes. Case S3 shows a strong improvement along the leading and trailing edge shear layers on the reference plane itself, as expected. A striking observation is the improvement along $z/c = 1.1$ for case S4 (Fig. \ref{fig:spatial_l1_single_0p52}). A plausible explanation for why both cases produce improvements on the off-data planes is their spanwise positioning relative to the symmetry boundary. The $z/c = 0.71$ plane is at an intermediate distance from the boundary, where the reference data and the symmetry condition together constrain the flow on both sides, allowing the correction to propagate to the neighboring planes. The $z/c = 0.52$ plane, being closest to the splitter plate, directly constrains the flow in a region where the baseline misfit is already small (Fig. \ref{fig:baseline_l1_norm}). Once this region is fixed, the governing equations provide sufficient constraint for the optimizer to infer the flow structure further from the boundary, explaining the improvement observed along $z/c = 1.1$.

\begin{figure*}[htbp]
    \centering
    \includegraphics[scale=0.17]{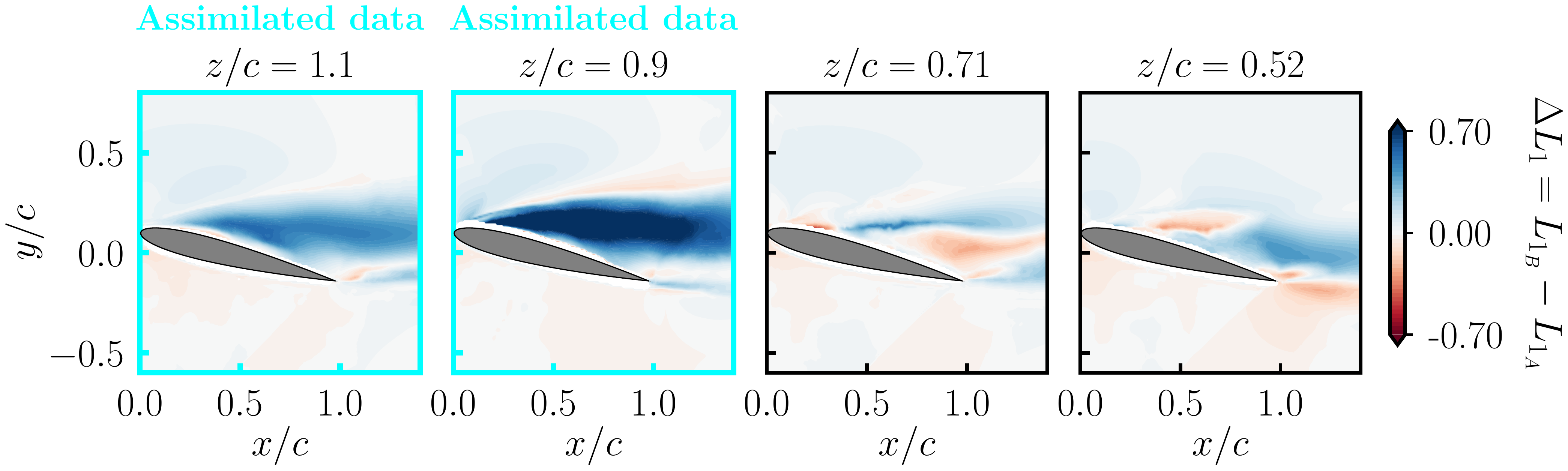}
    \caption{Spatial distribution of $\Delta L_1 = L_{1_B} - L_{1_A}$ across the four validation planes for the dual-plane case with reference data at $z/c = 1.1$ and $0.9$. Blue regions indicate improvement over the baseline and red regions indicate localized degradation. Colored bounding boxes denote the reference data planes (see Fig. \ref{fig:schematic} for the legend).}
    \label{fig:spatial_l1_improvement_dual}
\end{figure*}

The spatial distribution of $\Delta L_1$ for case D2 is presented in Fig. \ref{fig:spatial_l1_improvement_dual}. Both reference planes ($z/c = 1.1$ and $z/c = 0.9$) show a strong improvement, with the largest reductions in $L_1$ norm concentrated in the leading edge shear layer and the recirculation region, which are the areas of greatest baseline misfit (Fig. \ref{fig:baseline_l1_norm}). The improvement on $z/c = 0.9$ is particularly pronounced, which is expected given that this plane has the largest separation in the experimental data. The off-data planes ($z/c = 0.71$ and $z/c = 0.52$) remain largely unaffected, consistent with the trends observed in the waterfall plot and in the single-plane cases. This suggests that the influence of the assimilation is strongest in the spanwise region spanned by the reference data and diminishes towards the boundary where the baseline misfit is already small. While these results demonstrate that both the single-plane and dual-plane cases produce a measurable improvement in the mean velocity field, they do not reveal whether the assimilated flow has developed the 3D coherent structures characteristic of stall cells. A reduction in the misfit on individual planes does not guarantee that the spanwise organization of the flow or the surface flow topology is physically consistent with the presence of stall cells.

\subsection{Wake deficit and surface flow topology}
\label{subsection:wake_deficit_surface_topology}
To assess whether the assimilated flow fields exhibit the surface flow topology and wake structure associated with stall cells, the wake deficit, calculated as $1 - U_x/U_\infty$, and surface streamlines for the single and dual-plane cases are examined here. The streamwise velocity wake deficit at $70~\%$ chord ($x/c = 0.7$) is presented for both the single-plane and dual-plane cases. In addition, the spanwise velocity $U_z/U_\infty$ overlaid with in-plane streamlines ($YZ$ plane) is shown to indicate the extent of the spanwise flow along this plane. The choice of this streamwise location is motivated by the compact nature of the crossflow associated with stall cells. As demonstrated by \cite{neal2023three}, the strength of the streamwise vortices diminishes as the trailing edge is approached, indicating that they are highly localized. It is therefore more appropriate to sample the flow upstream of the trailing edge in order to capture these structures.

\begin{figure*}[htbp]
    \centering
        \begin{subfigure}{\textwidth}
        \centering
        \caption{}
        \includegraphics[scale=0.17]{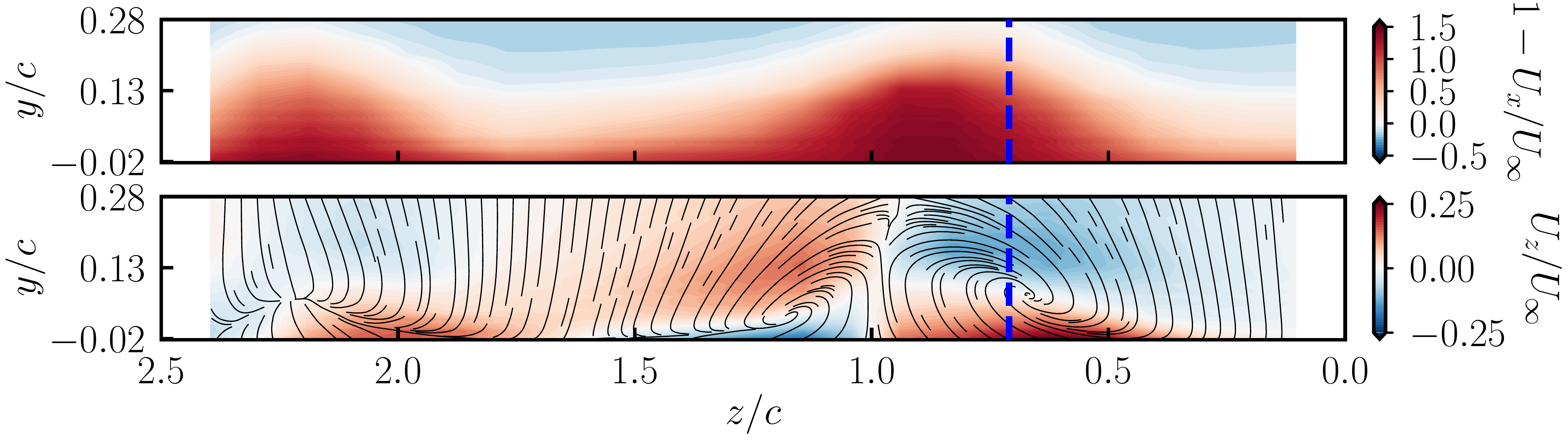}
        \label{fig:Combined_SuctionFlow_0p7_0p71}
    \end{subfigure}

    \vspace{0.5cm}
    
        \begin{subfigure}{\textwidth}
        \centering
        \caption{}
        \includegraphics[scale=0.17]{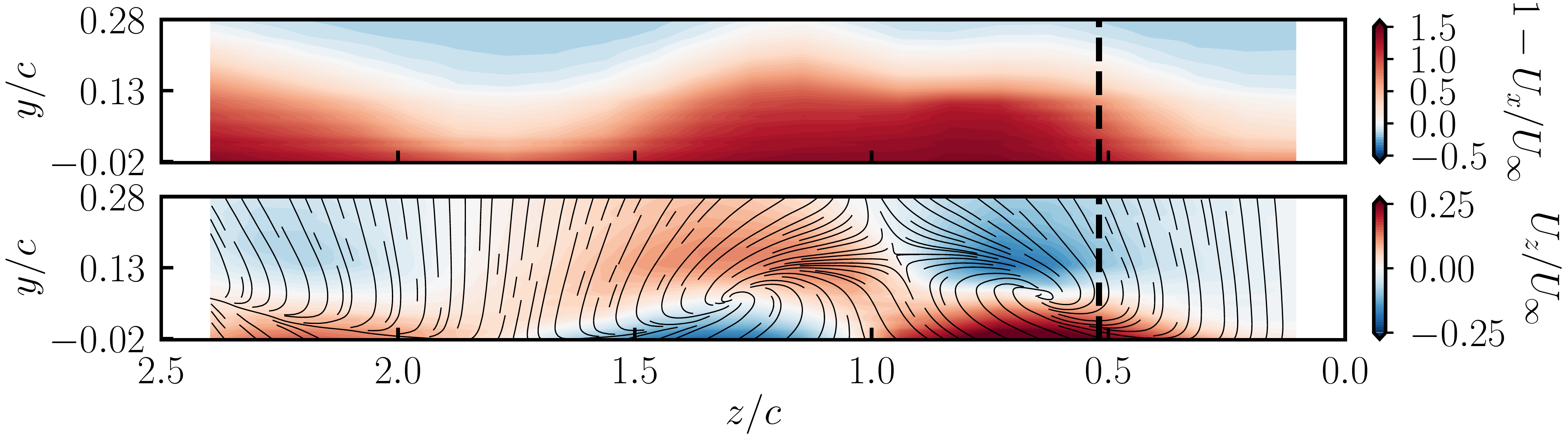}
        \label{fig:Combined_SuctionFlow_0p7_0p52}
    \end{subfigure}

    \caption{
    Assimilated suction-side flow characteristics at $x/c = 0.7$ for the single plane cases (a) S3 and (b) S4 (see Fig. \ref{fig:waterfall_absolute} for nomenclature). For each case, the upper subplot shows the streamwise velocity deficit ($1 - U_x/U_\infty$) and the lower subplot shows the spanwise velocity ($U_z/U_\infty$) with in-plane secondary flow streamlines. Dashed vertical lines indicate the spanwise locations of the reference data planes (see Fig. \ref{fig:schematic} for legend).}
    \label{fig:suction_flow_selected_single}
\end{figure*}

The mean streamwise velocity deficit and the spanwise velocity overlaid with in-plane streamlines for the single-plane assimilation cases are shown in Fig. \ref{fig:suction_flow_selected_single}. A clear region of strong recirculation is observed in both cases, corresponding to the presence of counter-rotating streamwise vortices indicated by the in-plane streamlines and the spanwise velocity contours. This is absent in the baseline (Fig. \ref{fig:suction_flow_baseline}). The presence of such streamwise vortices has been identified by \cite{neal2023three} as evidence for a stall cell. The spanwise locations of the vortex cores along the YZ plane, identified using the $\Gamma_1$ criterion \citep{graftieaux2001combining}, are reported in Table \ref{tab:core_locations}. In both single-plane cases, the inboard core (closer to the splitter plate at $z/c = 0$) is located at $z/c \approx 0.63$, confirming that this vortex core is anchored by the symmetry boundary regardless of where the reference data are provided. The outboard core, however, differs between the two cases. It is located at $z/c \approx 1.15$ for S3 and at $z/c \approx 1.35$ for S4, resulting in core-to-core distances of $0.52c$ and $0.73c$, respectively. Case S3, being further from the boundary, can work in conjunction with the symmetry condition to constrain the flow on both sides of the reference plane, producing a more compact reconstruction of the stall cell structure. Case S4, where the reference data are closest to the boundary, produces a more elongated structure as the optimizer has more freedom on the outboard side where no constraint is provided.

\begin{table}[H]
  \centering
  \caption{The spanwise locations of the counter-rotating vortex cores for Fig. \ref{fig:suction_flow_selected_single} and Fig. \ref{fig:suction_flow_selected_dual}, identified using the $\Gamma_1$ criterion \citep{graftieaux2001combining} at $x/c = 0.7$}
  \renewcommand{\arraystretch}{1.5}
  \begin{tabular}{lccc}
    \hline\hline
    Case & Inboard core ($z/c$) & Outboard core ($z/c$) & Core-to-core distance ($\Delta z/c$) \\
    \hline
     S3 & 0.63 & 1.15 & 0.52 \\
   \hline
     S4 & 0.63 & 1.35 & 0.73 \\
    \hline
     D2 & 0.52 & 1.15 & 0.63 \\
    \hline\hline
  \end{tabular}
  \label{tab:core_locations}
\end{table}

\begin{figure*}[htbp]
    \centering
        \includegraphics[scale=0.17]{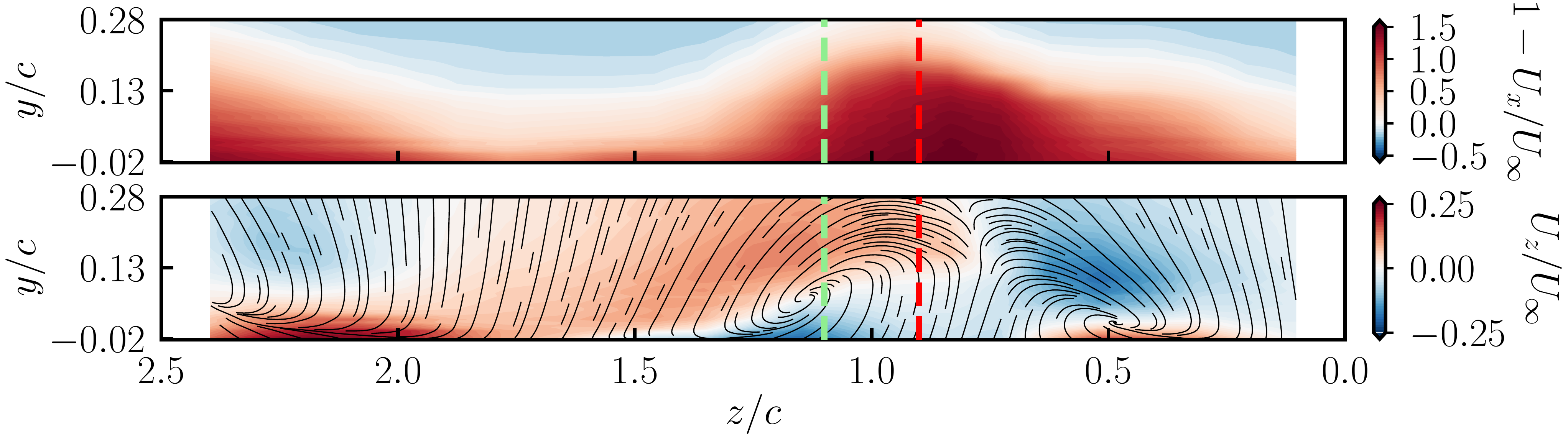}
    \caption{Assimilated suction-side flow characteristics at $x/c = 0.7$ for the dual-plane case with reference data at $z/c = 1.1$ and $0.9$. The upper subplot shows the streamwise velocity ($U_x/U_\infty$) and the lower subplot shows the spanwise velocity ($U_z/U_\infty$) with in-plane secondary flow streamlines. Dashed vertical lines indicate the spanwise locations of the reference data planes (see Fig. \ref{fig:schematic} for legend).}
    \label{fig:suction_flow_selected_dual}
\end{figure*}

The corresponding wake deficit and spanwise velocity for the dual-plane case are shown in Fig. \ref{fig:suction_flow_selected_dual}. The same features observed in the single-plane cases are present: a region of enhanced wake deficit accompanied by counter-rotating streamwise vortices visible in both the spanwise velocity contours and the in-plane streamlines. As reported in Table \ref{tab:core_locations}, the inboard core is located at $z/c \approx 0.52$ and the outboard core at $z/c \approx 1.15$, giving a core-to-core distance of $0.63c$. This value falls between the two single-plane cases ($0.52c$ for $z/c = 0.71$ and $0.73c$ for $z/c = 0.52$), suggesting that the dual-plane data provides a balanced constraint that prevents the structure from being either overly compact or overly elongated along the outboard direction. It is also notable that the inboard core in the dual-plane case is closer to the splitter plate ($z/c = 0.52$) than in either single-plane case ($z/c = 0.63$), which may reflect the additional constraint provided by the second reference plane at $z/c = 0.9$, drawing the structure slightly inboard. While the choice of reference data planes has a clear imprint on the reconstructed mean flow, particularly on the outboard core location and the overall compactness of the stall cell, the consistency of the inboard core across all three cases and the emergence of counter-rotating vortices in every instance demonstrates the capability of the variational DA framework to reconstruct a full 3D flow field from limited planar measurements. However, the wake deficit and in-plane velocity fields alone do not provide a definitive confirmation of the stall cell structure. For this, the surface flow pattern must be examined.

\begin{figure*}[htbp]
    \centering
    
    \begin{subfigure}{\textwidth}
        \centering
        \caption{}
        \includegraphics[scale=0.17]{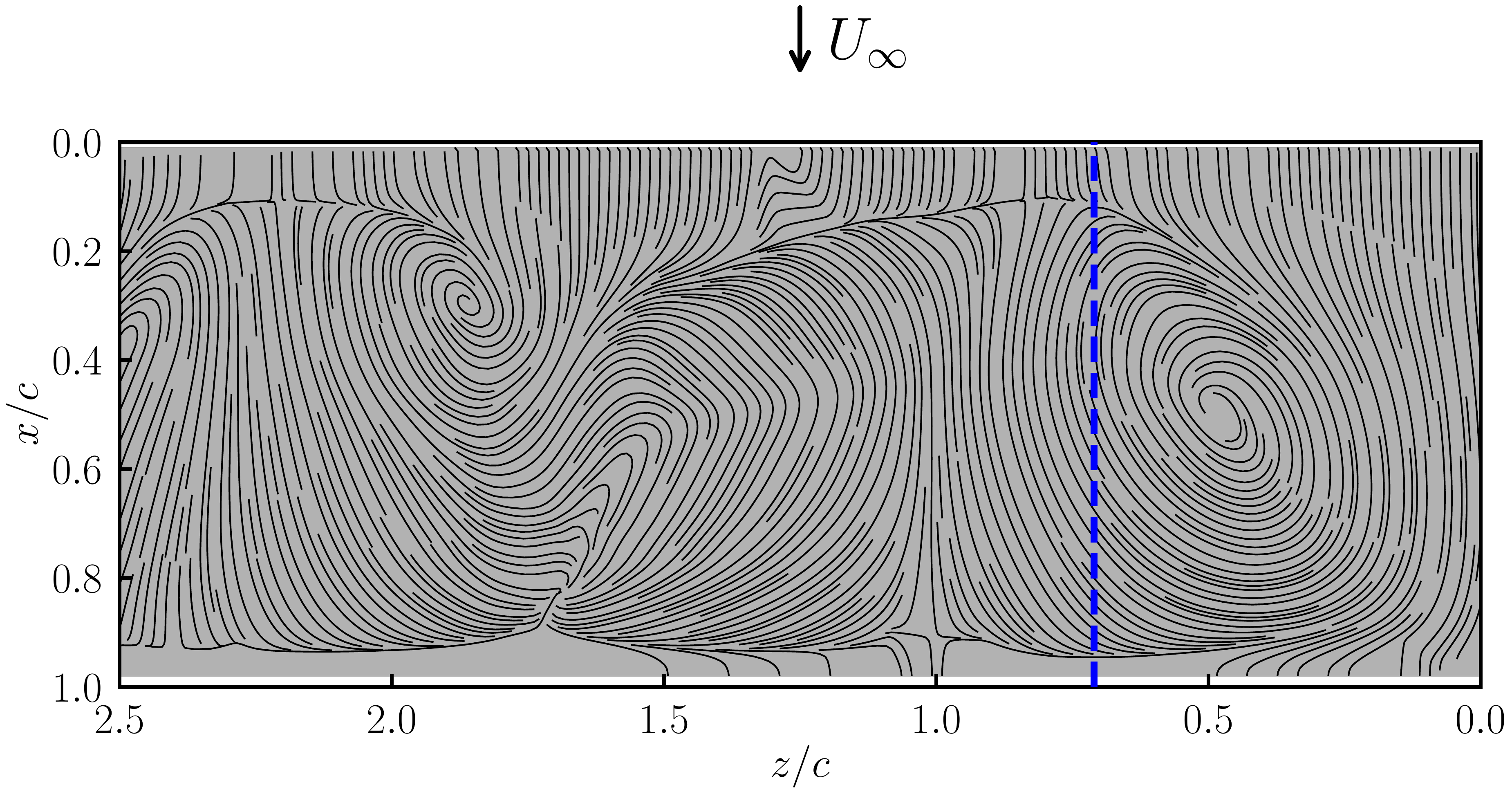}
        \label{fig:wss_single_0p71}
    \end{subfigure}
    
    \vspace{-0.5cm}
        \begin{subfigure}{\textwidth}
        \centering
        \caption{}
        \includegraphics[scale=0.17]{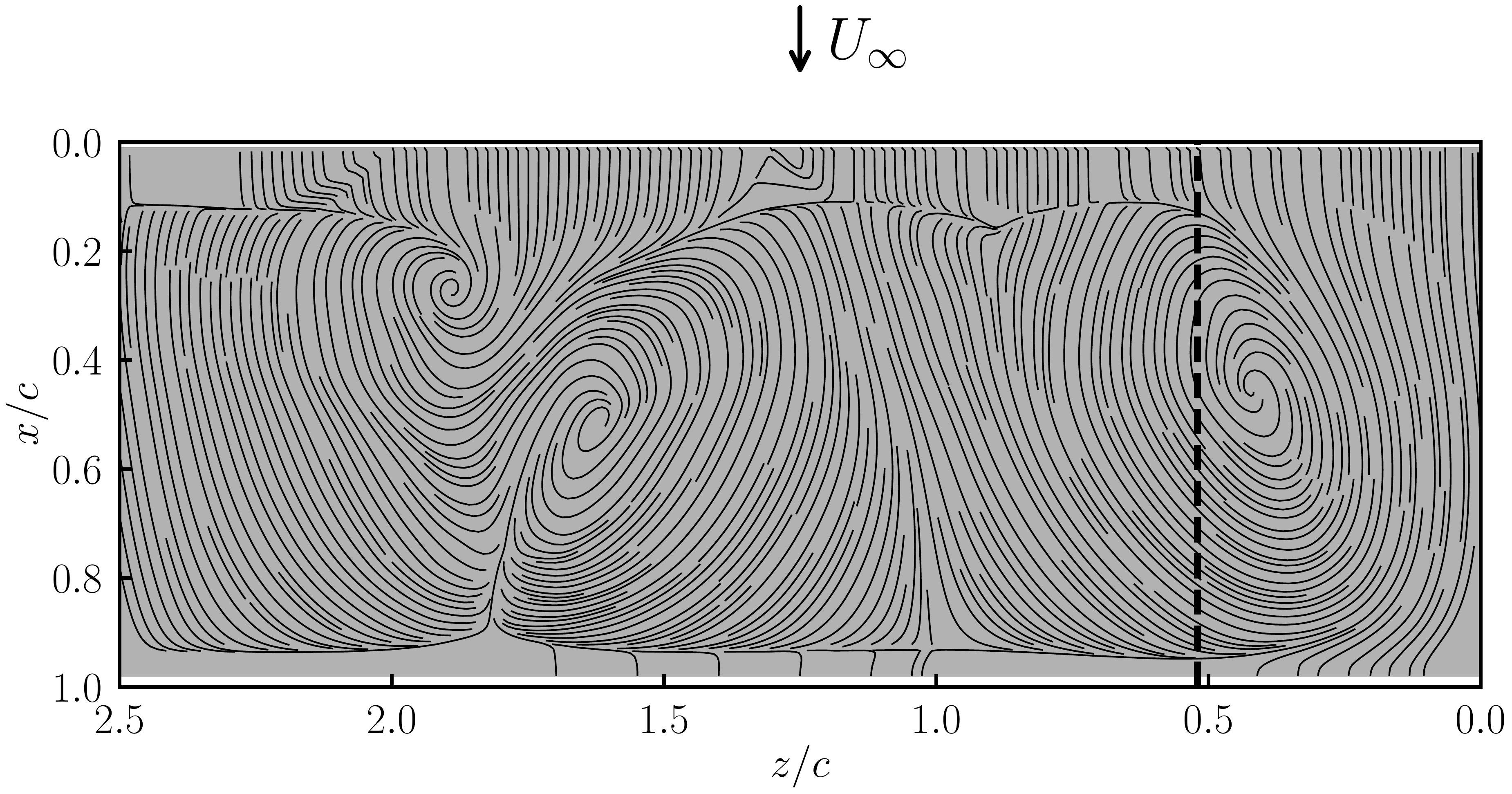}
        \label{fig:wss_single_0p52}
    \end{subfigure}

    \caption{Surface streamlines computed from the wall shear stress components ($\tau_x$, $\tau_z$) for the single-plane cases (a) S3 and (b) S4. Dashed vertical lines indicate the spanwise locations of the reference data planes (see Fig. \ref{fig:schematic} for the legend).}
    \label{fig:wss_single}
\end{figure*}

The defining feature of a stall cell, as documented across decades of experimental and computational studies, is the presence of counter-rotating vortices around focal points on the suction surface of the wing \citep{winklemann1980, yon1998study, manolesos2014experimental, neal2023three, liu2018numerical}. These focal points of the vortices, connected by a curved separation line, produce the characteristic ``owl's eye'' or ``mushroom-shaped'' pattern that is visible in oil-flow visualizations and in computed surface streamlines. Having established the presence of streamwise vortices at $x/c = 0.7$, the surface flow topology for the three selected cases is now examined to determine whether the assimilated flow fields reproduce this defining pattern. The surface streamlines computed from the wall shear stress components ($\tau_x$, $\tau_z$) for the two single-plane cases are shown in Fig. \ref{fig:wss_single}. Case S3 (Fig. \ref{fig:wss_single_0p71}) shows only one clearly formed focal point on the inboard side, with a weak and incomplete formation of the outboard focal point. Case S4 (Fig. \ref{fig:wss_single_0p52}) shows two well-defined focal points on both the inboard and outboard sides. Both cases suggest the presence of approximately $1.5$ stall cells within the computational domain. The spanwise wavelength of the stall cell, as seen by the surface flow topology, is $\lambda_z \approx 1.4c$ for case S3 and $\lambda_z \approx 1.6c$ for case S4. The larger recirculation region for case S4 is consistent with the elongated region of strong recirculation towards the outboard side observed in Fig. \ref{fig:suction_flow_selected_single} and reflects the inability of a single plane of data close to the boundary to adequately constrain the outboard extent of the stall cell. Case S3, being further from the boundary, produces a more compact recirculation region.

\begin{figure*}[htbp]
    \centering
    \includegraphics[scale=0.17]{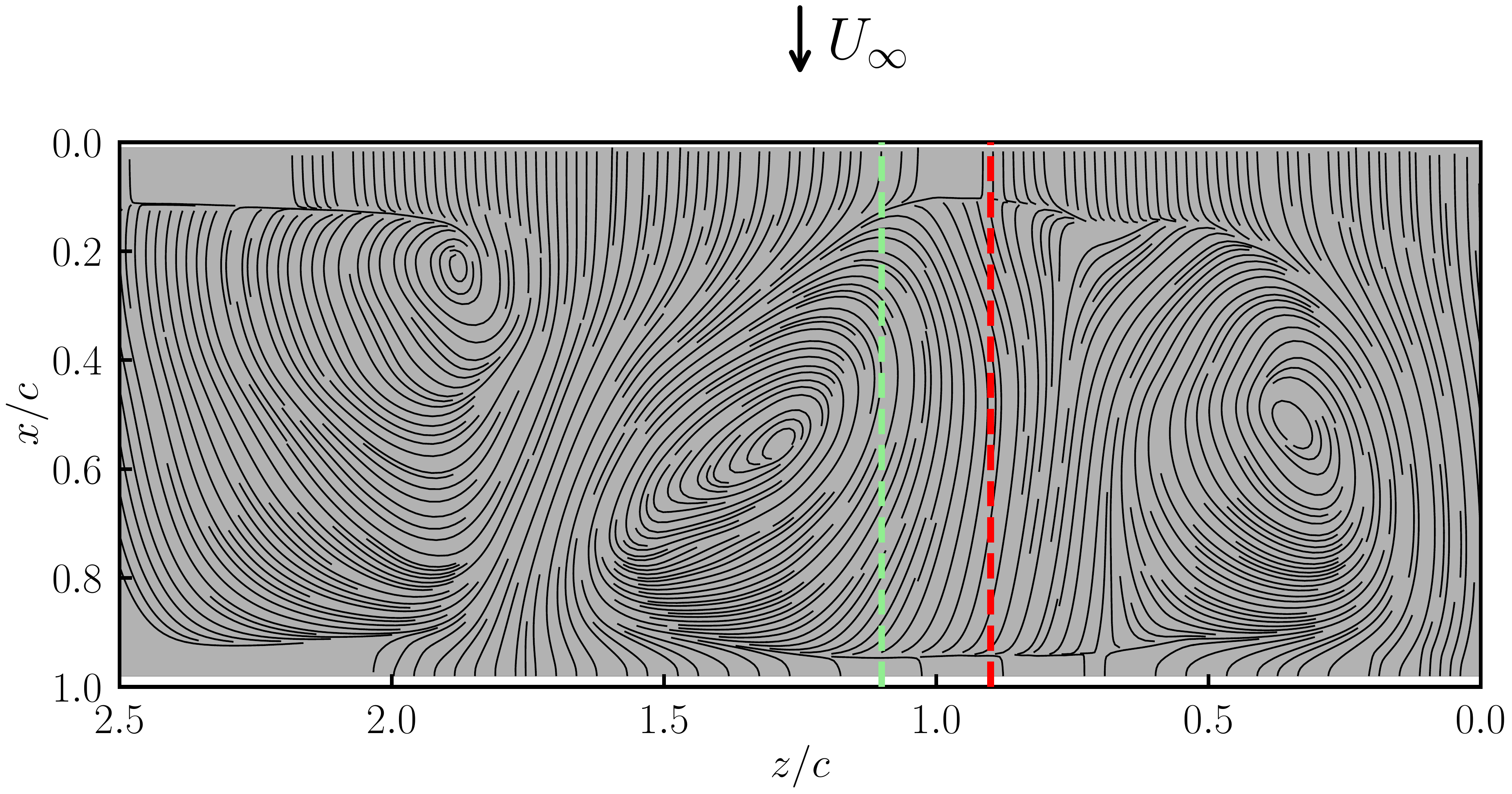}
    \caption{Surface streamlines computed from the wall shear stress components ($\tau_x$, $\tau_z$) for the dual-plane case with reference data at $z/c = 1.1$ and $0.9$. Dashed vertical lines indicate the spanwise locations of the reference data planes (see Fig. \ref{fig:schematic} for the legend).}
    \label{fig:wss_dual}
\end{figure*}
%% Use \subsection commands to start a subsection.

The surface flow topology for case D2 is shown in Fig. \ref{fig:wss_dual}. A clearly defined stall cell with well-formed focal points on both the inboard and outboard sides is observed. The spanwise wavelength of the stall cell is estimated as $\lambda_z \approx 1.3c$, which is consistent with the wavelengths reported in the literature for comparable configurations: $\lambda_z \approx 1.25c$ for a NACA 4412 at $Re_c = 350{,}000$ \citep{sarras2024linear} and $\lambda_z \approx 1.25c$ for a NACA 0012 at $Re_c = 10^6$ \citep{liu2018numerical}. This is a more compact structure than both single-plane cases, which produced wavelengths of $1.4c$ (S3) and $1.6c$ (S4). The presence of two reference planes that are close together in physical space but markedly different in their mean velocity fields provides the optimizer with direct information about the rapid change in separation extent along the span. It is also notable that the inboard focal point forms at approximately the same spanwise location ($z/c \approx 0.5$ to $0.6$) as in cases S3 and S4, further confirming that the symmetry boundary anchors this end of the stall cell independently of the data configuration.

The role of the continuity constraint in enabling the full-field reconstruction is now discussed. It has been shown that planar PIV data of stalled airfoils are not inherently divergence-free due to the presence of a non-negligible out-of-plane mean velocity component \citep{cadambi2026three}. In a 3D RANS simulation, the continuity equation $\partial U_x / \partial x + \partial U_y / \partial y + \partial U_z / \partial z = 0$ must be satisfied at every iteration of the optimizer. When the correction field $\beta$ drives the streamwise and wall-normal velocity components toward the experimental values on the reference planes, the continuity equation forces a spanwise velocity $U_z$ to develop that is consistent with the imposed spanwise gradients. This component is not prescribed by the data — it emerges from the governing equations. Furthermore, since the correction operates within the Boussinesq framework, the permissible changes to the Reynolds stress tensor are constrained to those representable by an eddy viscosity. The combination of the continuity constraint and the Boussinesq closure restricts the solution space such that the presence of sparse experimental data is sufficient to guide the optimizer toward a physically meaningful 3D flow state that the baseline model alone cannot reach.

\begin{figure*}[htbp]
    \centering
    \begin{subfigure}{0.48\textwidth}
        \centering
        \caption{}
        \includegraphics[width=\textwidth]{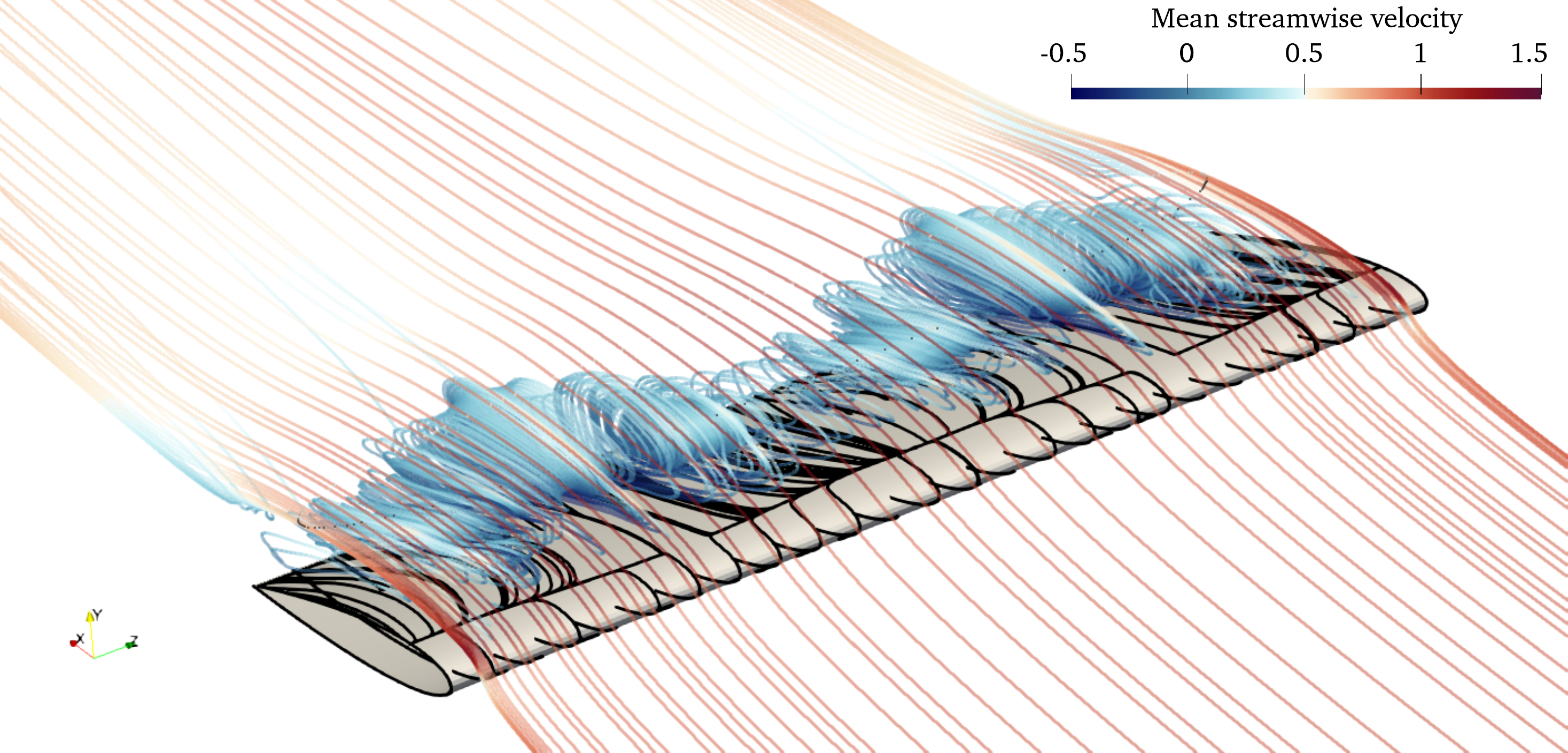}
        \label{fig:bsl}
    \end{subfigure}
    \hfill
    \begin{subfigure}{0.48\textwidth}
        \centering
        \caption{}
        \includegraphics[width=\textwidth]{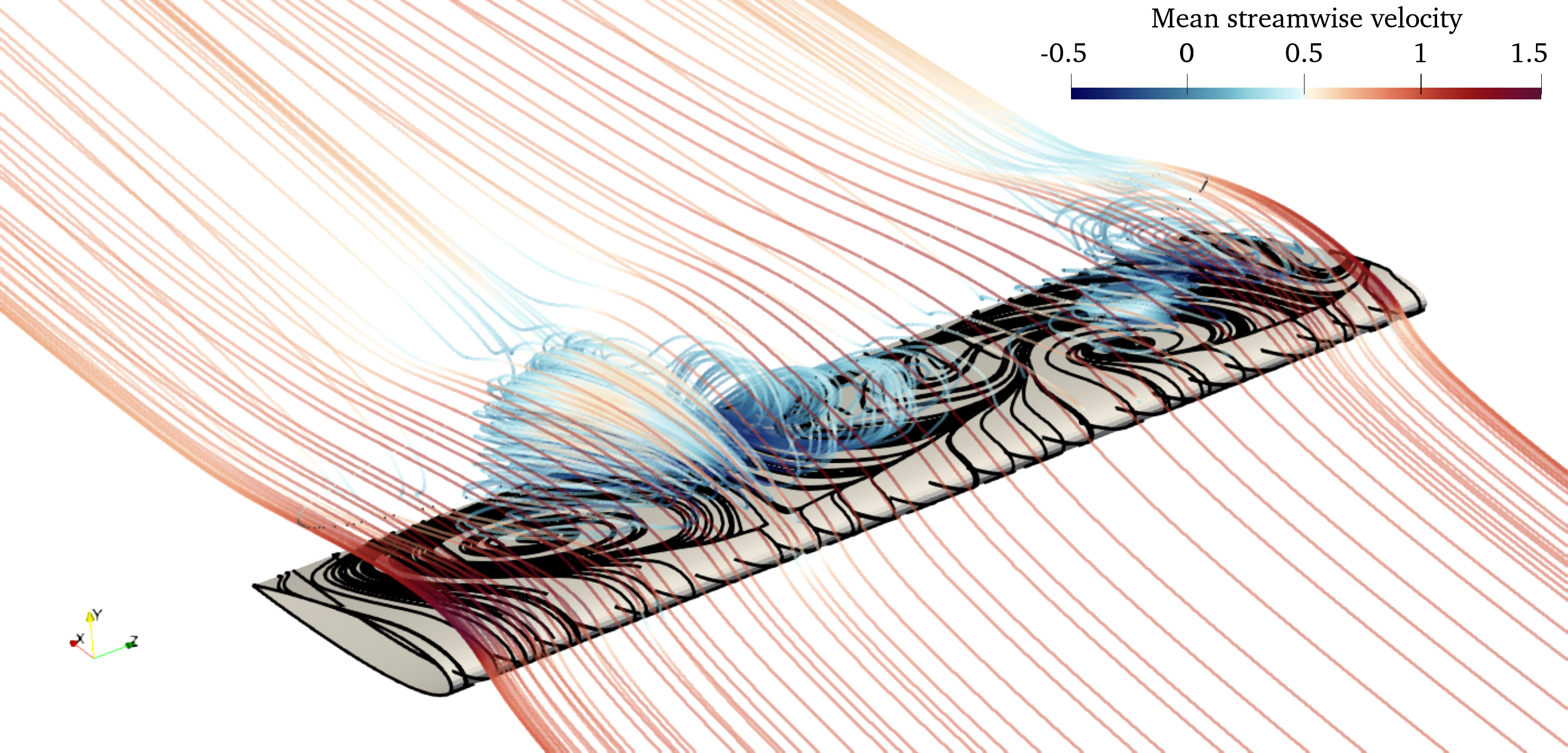}
        \label{fig:p3}
    \end{subfigure}
    
    \vspace{6pt}
    
    \begin{subfigure}{0.48\textwidth}
        \centering
        \caption{}
        \includegraphics[width=\textwidth]{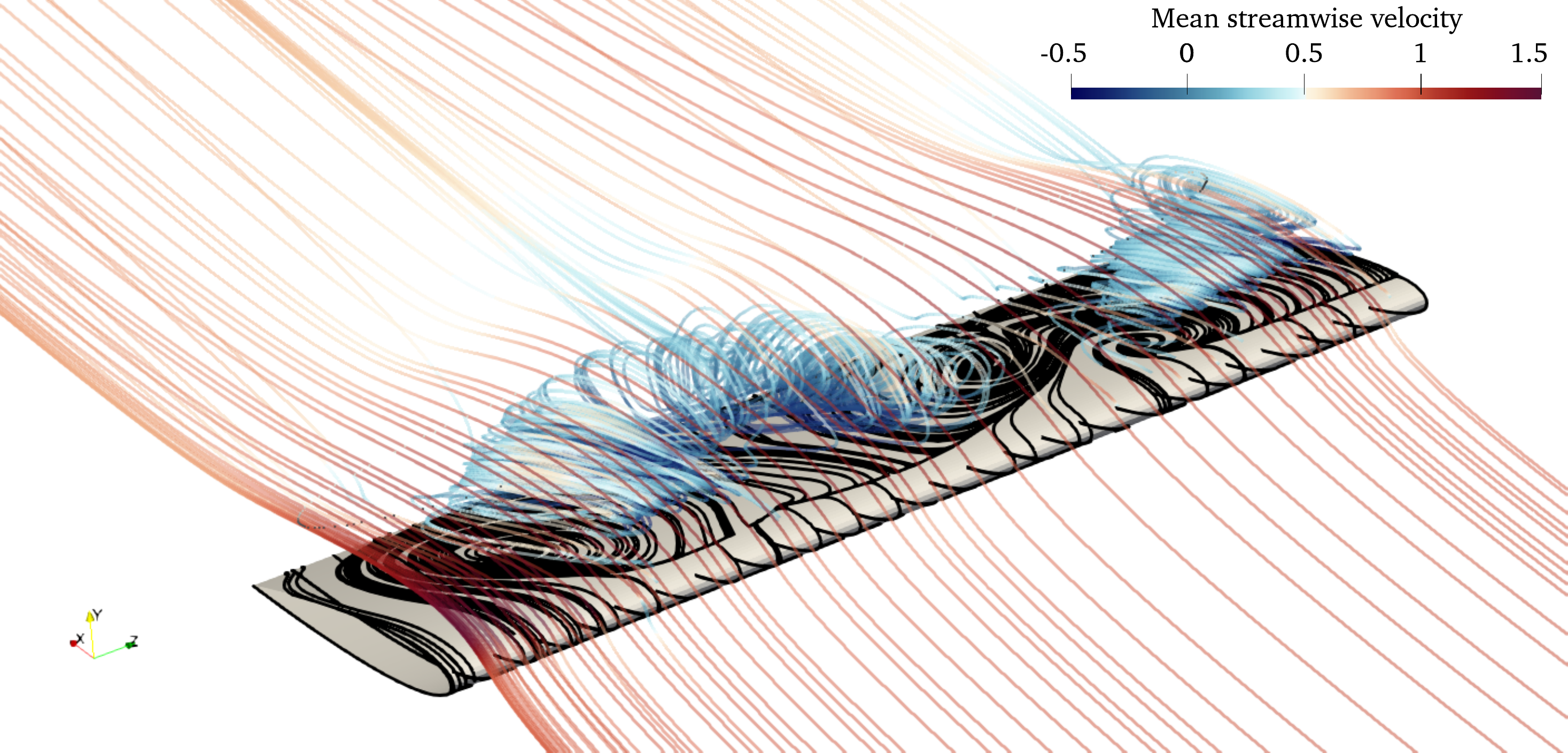}
        \label{fig:p2}
    \end{subfigure}
    \hfill
    \begin{subfigure}{0.48\textwidth}
        \centering
        \caption{}
        \includegraphics[width=\textwidth]{figures/figure_18d.pdf}
        \label{fig:p1p4}
    \end{subfigure}
    
  \caption{Volumetric streamlines seeded from a spanwise-elongated elliptical streamtube on the suction side, colored by the mean streamwise velocity $U_x/U_\infty$, with surface streamlines computed from wall shear stress components $\tau_x$ and $\tau_z$. (a) Baseline, (b) S3, (c) S4, (d) D2.}
    \label{fig:blob_contours}
\end{figure*}

\section{Conclusion}
\label{section:conclusion}
We perform planar PIV experiments on a NACA 0012 wing at $Re_c \approx 450{,}000$ and $\alpha = 14^\circ$ along four spanwise planes separated by $\Delta z\approx 0.2c$. The flow on the four planes shows characteristics of spanwise-organized coherent structures called stall cells, both in the mean velocity fields, and in the dynamics as captured by POD. The baseline SA model fails to produce stall cells at these conditions despite exhibiting weak three-dimensionality. To reconstruct the full 3D mean flow field, 3DVar DA is performed using the field inversion framework with sparse two-component planar PIV data. By correcting the production term of the turbulence transport equation using data from at most two of the four available spanwise planes, the assimilated flow fields recover the defining features of stall cells: counter-rotating vortices around focal points on the surface, increase in the magnitude of the spanwise component of mean velocity (reaching $30~\%$ of the freestream value), and spanwise wavelengths of the vortical structures on the surface of the wing in the range $\lambda_z \approx 1.3c$ to $1.6c$, consistent with values reported in the literature \citep{sarras2024linear, liu2018numerical, gross2015criterion}. The reconstruction is validated on planes not used in the assimilation, confirming that the governing equations, in particular the continuity constraint, propagate the corrections well beyond the spatial extent of the reference data.

A key finding is that the inboard focal point of the stall cell forms at a consistent spanwise location ($z/c \approx 0.5$ to $0.6$) across all assimilation cases, regardless of the number or placement of reference planes. This demonstrates the complementary roles of the experimental data and the computational boundary conditions in shaping the reconstructed flow. The reference data introduces the three-dimensionality that the baseline model lacks, while the symmetry boundary constrains one end of the stall cell. The continuity equation develops the spanwise velocity field that organizes the full 3D structure. The 3D structure of the reconstructed stall cell is further illustrated in Fig. \ref{fig:blob_contours}, which shows volumetric streamlines colored by the mean streamwise velocity alongside surface streamlines computed from $\tau_x$ and $\tau_z$. For case D2 (Fig.~\ref{fig:p1p4}), the streamlines within the recirculation region form a helical structure stretched into an inverted U-shape, with two counter-rotating legs rising from the surface focal points and connecting above the suction surface. This arch-like topology is consistent with the arch vortex identified by \cite{neal2023three} in their experimental study of stall cells on a cantilevered wing, where the two surface foci were found to be connected by a single coherent vortical structure that lifts off the surface and extends into the wake. A similar structure is observed for case S4 (Fig. \ref{fig:p2}), though the arch is more elongated in the spanwise direction, consistent with the larger stall cell wavelength produced by this case. For case S3 (Fig. \ref{fig:p3}), the helical structure is less well-defined, which is consistent with the incomplete formation of the outboard focal point observed in the surface streamlines. The baseline (Fig. \ref{fig:bsl}) shows no such organized structure, with the streamlines remaining largely two-dimensional within the recirculation region. The progression from the baseline to the assimilated cases confirms that DA not only recovers the correct surface flow topology but also reconstructs the full 3D vortical architecture of the stall cell.

Several directions for future work are envisaged. Localized adjoint methods could reduce the substantial memory requirements of 3D adjoint-based assimilation and enable the use of finer meshes with a resolved viscous sublayer. The optimized $\beta$ fields also provide a dataset for symbolic regression approaches that could distill the corrections into generalized algebraic models for stall cell flows, completing the second step of the FIML framework.

\begin{appendices}
\section{POD computation and cross-plane correlation}
\label{appendix:pod}

Proper orthogonal decomposition is applied independently to the instantaneous velocity snapshots on each of the four spanwise measurement planes. For a given plane, the $N = 1000$ snapshots of the two-component velocity field ($U_x$, $U_y$) are arranged into a data matrix $\mathbf{D} \in \mathbb{R}^{M \times N}$, where each column contains the flattened and concatenated velocity components at all valid spatial points (i.e.\ excluding masked regions near the airfoil surface). The temporal mean is subtracted from each row to form the fluctuation matrix $\mathbf{D}' = \mathbf{D} - \bar{\mathbf{D}}$. The singular value decomposition $\mathbf{D}' = \boldsymbol{\Phi} \mathbf{S} \mathbf{V}^T$ is then computed, where the columns of $\boldsymbol{\Phi}$ are the spatial modes, the diagonal entries of $\mathbf{S}$ are the singular values, and the columns of $\mathbf{V}$ are the temporal coefficients. The energy fraction of mode $n$ is given by $\lambda_n / \sum_n \lambda_n$, where $\lambda_n = s_n^2$ and $s_n$ is the $n$-th singular value.

To assess the spanwise coherence of the POD modes across different measurement planes, a cross-plane correlation analysis is performed. Since the four PIV planes have slightly different fields of view due to variations in the optical setup, the spatial modes are first interpolated onto a common grid. Calibrated coordinate translations, determined by aligning the mean velocity contours of each plane to a reference plane ($z/c = 1.1$), are applied before interpolation. Only the spatial region common to all four planes after translation is retained, ensuring that the correlation is computed over an identical domain for every plane pair. For each pair of planes ($p$, $q$) and each pair of mode indices ($m$, $n$), the Pearson correlation coefficient is computed between the corresponding spatial mode vectors as
\begin{align}
\rho_{mn}^{pq} = \frac{\sum_i (\phi_m^p(i) - \bar{\phi}_m^p)(\phi_n^q(i) - \bar{\phi}_n^q)}{\sqrt{\sum_i (\phi_m^p(i) - \bar{\phi}_m^p)^2} \sqrt{\sum_i (\phi_n^q(i) - \bar{\phi}_n^q)^2}},
\end{align}
where $\phi_m^p(i)$ denotes the $i$-th spatial point of mode $m$ on plane $p$, and the summation runs over all valid points in the common domain. Both velocity components are concatenated into a single vector before computing the correlation. The absolute value $|\rho_{mn}^{pq}|$ is reported, since modes of opposite sign represent the same spatial structure. This procedure yields a $10 \times 10$ correlation matrix for each of the six unique plane pairs, as presented in Fig. \ref{fig:pod_correlation}.
\end{appendices}

\section*{Acknowledgements}
The authors acknowledge the use of the IRIDIS5 and IRIDIS6 High Performance Computing Facility, and associated support services at the University of Southampton, in the completion of this work.

\section*{CRediT authorship contribution statement}
\textbf{Uttam Cadambi Padmanaban} --  Methodology, developing the software, Formal analysis, Data Curation, Writing - Original Draft. \textbf{Craig Thompson} -- Investigation, Data Curation, Writing - Original Draft . \textbf{Bharathram Ganapathisubramani} \& \textbf{Sean Symon} -- Conceptualization, Writing - Review \& Editing, Supervision, Project administration, Funding acquisition.

\section*{Funding}
We gratefully acknowledge funding from EPSRC (Grant Ref: EP/W009935/1) and the School of Engineering at University of Southampton for CT's and UCP's PhD studentships.

\section*{Declaration of competing interest}
The authors declare that they have no known competing financial interests or personal relationships that could have appeared to influence the work reported in this paper.

\bibliographystyle{elsarticle-harv} 
\bibliography{cas-refs}



%\begin{thebibliography}{00}


%\bibitem[Lamport(1994)]{lamport94}
%  \textit{\LaTeX: a document preparation system},
%  Addison Wesley, Massachusetts,

%\end{thebibliography}
\end{document}